\documentclass{article}
\usepackage{amsmath}
\usepackage{hyperref}
\usepackage{graphicx}
\usepackage{xcolor}
\usepackage{tikz}
\usepackage{hanging}
\usepackage{float}
\usepackage[margin=1cm,font=small]{caption}
\usepackage{wrapfig}
\usepackage{placeins}
\usepackage{setspace}
\newif\ifforReview
\newif\iffiguresInText
\forReviewfalse        
\figuresInTexttrue     

\title{Massive Genealogies Distinguish Frontier from Steady-State Internal Migration}
\ifforReview
\else
    \author{Robin W. Spencer\footnote{The author is a retired biochemist and data analyst, email: spencerrw@alum.mit.edu}\space \space and Samuel M. Otterstrom\footnote{Department of Geography, 690 KMBL, Brigham Young University, Provo, UT 84602, e-mail: otterstrom@byu.edu }}
\fi

\begin{document}
\sloppy   
\date{\vspace{-5ex}}  
\maketitle

\begin{abstract}Recent studies of human migration have focused on modern issues of international economics, politics, urbanization, or commuting.  Here we make use of very large anonymized genealogies which offer quantitative metrics and models before census data became available.  In European and North American data from 1400 to 1950 we find two distinct patterns of lifetime migration. The steady-state pattern shows a universal power-law distribution of migration distance; by its early appearance it cannot be dependent on post-industrial technology.  The frontier pattern, in contrast, is not scale-free with its much longer average distances.  All migration distances are well fit by a three parameter model; the temporal and geographic patterns of the fitted parameters give new insight to American internal expansion 1620-1950. Frontier migration is also highly directional and asymmetric; gravity models do not apply.  The American frontier pattern arose from the colonial-era steady-state within a generation, plateaued for three generations, then returned to a more mobile steady-state, a sequence paralleled by the Steppe migrations that brought the Bronze Age to Neolithic Europe. The transient frontier pattern is enabled by large-scale technological or numeric imbalance and geographic opportunity; when these forces abate, a new steady-state begins.
\end{abstract}

\textsc{keywords} \small\textsf{migration, frontier,distance distribution, genealogies, modeling}

\section*{Introduction}
    Humans must have a way to physically survive in order to live in a particular place. Food and shelter are paramount, but the processes by which people have supported themselves across the diverse habitats of the world have varied in countless ways over the centuries. At times necessity or opportunity has resulted in migrations of individuals and families from one locale to another. Sometimes their destination can be near their current home and other times it may be hundreds or thousands of miles away. The patterns and processes of human migration have been studied by scholars around the globe focusing on various historical periods and places. Their studies have resulted in a nearly endless array of articles and books that aim to discover the main patterns and processes of life course migration and even multi-generational migration. In this paper we propose a model of migration that combines a limited set of variables  that describe the timing and geographical spread of migrants historically. We use a rich genealogical dataset (Kaplanis et al 2018) to illustrate the applicability of our model to different places and during different periods.

Analyzing migration patterns has often been limited by the amount and quality of population data. Census records and other government compilations, such as birth or death records, have been used extensively, but they are limited in that they are just one snapshot of people at one time. However, in some censuses, for example, the birth location of an individual and their parents are included (like the 1880 US Census) along with a location for where the person resided at that time, which therefore give a two-generation data picture of migration at that point. US censuses can also have some other characteristics that increase their value, such as including socio-economic variables like income, education, and occupation and other variables such as race and spoken language depending on what is on the census forms. As such, using censuses and other government vital records for data sources to study migration within the US and other nations, has been the norm for many years.

Scholars have explored the potential of the linking multiple sets of records together, such as subsequent censuses. This has been a valuable approach, but it also has limitations. Some of these are underreporting of certain populations, mistakes in the records caused by the census takers and the people being enumerated, and other issues such as missing people between censuses (Knights 1969; Knights 1991a; Anderson 1988; Anderson and Fienberg 1997; Anderson 2019; Winchester 1970; Hershberg et al 1976; Markoff and Shapiro 1973; Ruggles et al 2018; Pouyez et al 1983). Notwithstanding these challenges, researchers have used record linking and comparing among censuses, to help them study topics such as historical migration (Thernstrom and Knights 1970; Hudson 1988; Steckel 1988), economic and social mobility over time (Thernstrom 1964; Thernstrom 1973), growth in ethnic neighborhoods (Chudacoff 1973; Conzen 1979), cultural diffusion (Waters et al 2007; Arreola and Hartwell 2014; Ueda 2017), and community population structure and urban history (Chudacoff 1996; Knights 1971; Knights 1991b; Conzen 1981).

In our research, we do not use linked census records, but instead rely on a high-quality genealogical dataset. For years scholars have recognized how valuable genealogical data could be for population research (e.g., Mineau et al 1989 ; Adams et al 2002). Until recently, the small sizes of genealogical datasets and issues of data quality have been challenging and have brought up questions concerning how representative the data are in terms of the full population.  However, genealogical data have increasingly been used in population and migration research as the size and quality of genealogy datasets have grown and technology improved (e.g., Otterstrom and Bunker 2013; Koylu et al 2014; Han et al 2017; Kaplanis et al 2018; Koylu and Kasakoff 2022). 

Indeed, in this paper, using genealogical data, we focus on internal migration, and in particular how the distributional properties of a single measurement — lifetime migration distance as recorded by birth and death dates and locations in each person’s genealogic profile — can be used to model general migration processes.   These distance properties, together with directionality and symmetry measures derived from the same profiles, define a unique pattern that coincides with the American westward expansion era.  Our use of genealogical data is key: its precision — individuals’ specific dates and locations at a 10 km scale — lets us see complete distributions, and so build a comprehensive distance model.  Its quantity permits analysis of temporal and geographic subsets which, by correlation to known history, lets us interpret the numeric results. Censuses alone, and even linked censuses, simply do not have all of that information in a readily usable form.

\section*{Inferences from Pre-Industrial Data}

Cell-phone, social media, and financial transaction datasets have transformed 21st century demography (Deville et al. 2014, Blazquez, D., \& Domenech, J. 2018, Barbosa et al 2018) and unsurprisingly these observations are often interpreted in the context of a modern technological society.   Large crowd-sourced genealogies are the inverse: they are generally poor resources for events after about 1950, since privacy and preferences generally exclude living individuals and their immediate relatives. But by reaching back well before the Industrial Revolution, genealogies can reveal patterns that pre-date industrial and urban transitions.  These patterns cannot have origins in modern communications, technology, or transportation, but must reflect something less technological and more universal about human behavior.  If an early pattern persists in recent datasets, then parsimony suggests that the pre-industrial rationale should prevail.  Before discussing human migration, we briefly note such a case for the oldest-known distribution in demography.

\section*{Demographic Parallel in City Sizes}

The long-tailed distribution of city sizes was first noted by Auerbach (1913).  As shown in Figure 1, the pattern spans at least two millennia from the cities of the Roman Empire, to early medieval England, to the modern era.  Most of the large literature devoted to this pattern focuses on recent data and seeks explanations in classical economic terms, in which well-informed, rational actors move to improve their economic well-being (Gabaix \& Ioannides 2004, Batty 2013).

\iffiguresInText
    \begin{figure}[H]
        \centering
        \includegraphics[width=10cm]{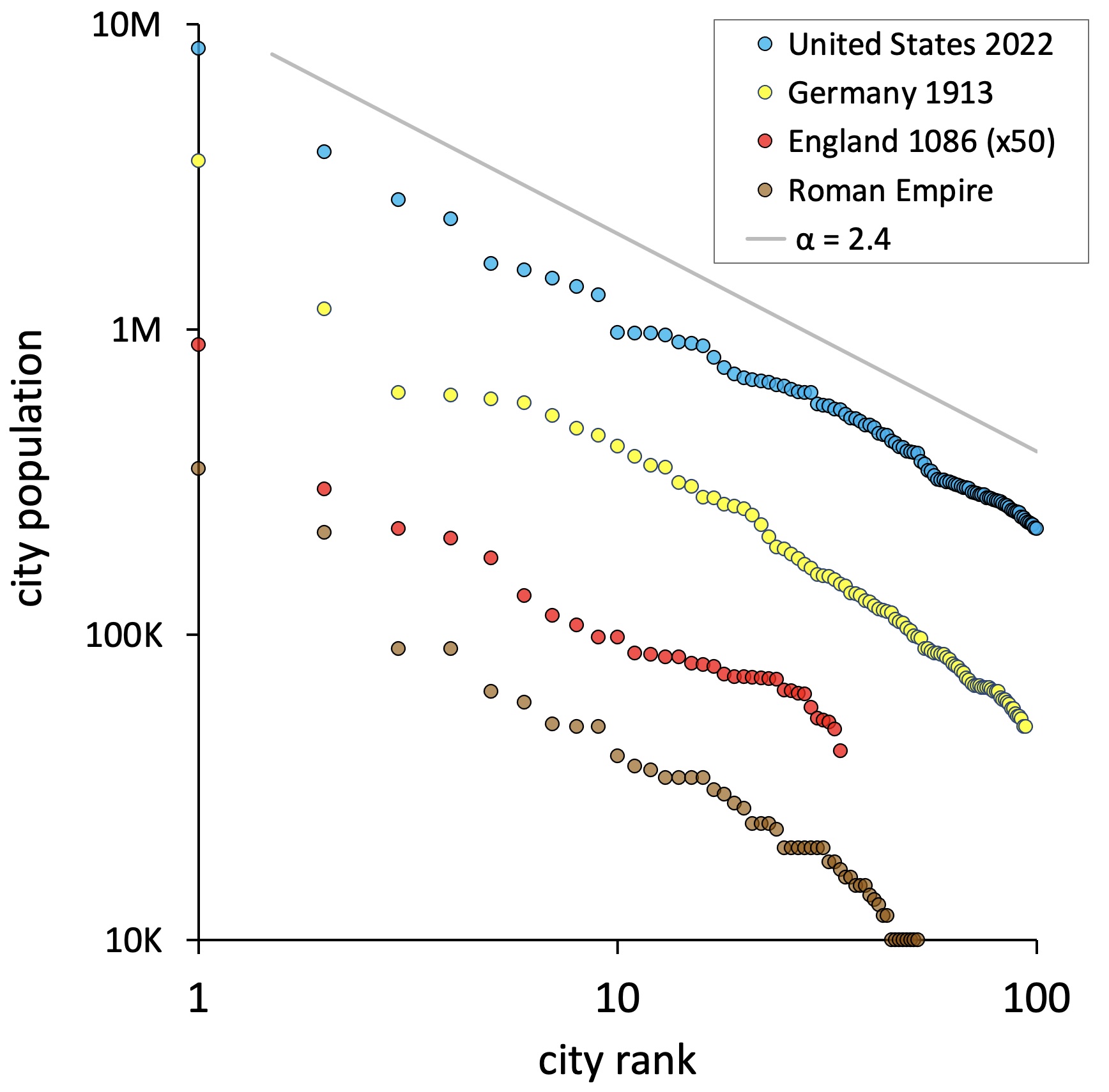}
        \caption{Examples of rank-size distributions for city sizes, ancient to modern.  The gray line is a power-law function with exponent -2.4, which is typical of these distributions.  See Spencer (2024) and \url{http://scaledinnovation.com/gg/cityGrowth.html} for datasets, interactive curve fitting, and source code.}
        \color{gray}\rule{10cm}{0.5pt}
    \end{figure}
\fi

The antiquity of the pattern argues for a different and universal alternative, namely that the power law distribution is simply the result of children making the inevitable stay-versus-leave-home decision as they reach adulthood (Spencer 2024). If they stay (historically the majority decision), their own headcount attaches to their birth location, which is the preferential attachment that drives the power law, and if they leave, they seed (or contribute to) another community.  This model is self-initializing, unlike a city-to-city migration model, and so it is timeless and applicable from the earliest fixed settlements.  This exponent of the power law is a simple function of the stay-vs-leave probability and always greater than two, as observed.

\section*{A Universal Pattern of Migration}

Similar to city sizes, our principal dataset (Kaplanis et al. 2018) reveals a demographic distribution that is centuries older than usually recognized or rationalized.  In this case the relevant metric is not city size, but lifetime migration distance as measured by genealogic birth and death locations recorded as latitude and longitude.  Figure 2a shows the distribution of this distance for 1.1 million Kaplanis profiles plus 9.8 million US census profiles with a result that is shared across continents and centuries: migration distances follows a power law with exponent -1.  The probability that a person has a given lifetime migration distance is inversely proportional to that distance.  The same data (with additional time-slices) are shown in Figure 2b as normalized cumulative distributions of those who moved at least 1 km, where normalization makes the consistency across continents and centuries more evident.  Since the integral of 1/x is ln(x), these datasets should be predominantly linear in the cumulative semilog presentation of Fig 2b, as observed.

There are three systematic deviations from a power law (or linearity in Fig 2b); two are simple and technical and one reflects exceptional events.  The first deviation occurs at small distances: below about 10 km for the genealogies and 100 km for the 2010 census study, observed counts fall below the 1/r distribution.  This reflects limited distance resolution: users of online genealogies typically enter an ancestor’s town or county but not a street address, which collapses short moves to zero.  Other sources have coarser resolution, for example Ravenstein (1885) defines migration as between English counties and the 2010 US census study as between commuting zones (Sprung-Keyser et al 2022).  Cumulative distributions avoid the issue and allow all profiles to be tallied, and with a linear vertical axis they focus attention evenly across the population.  We note that much of the literature calculates “percent migration” based on an arbitrary distance, such as 10 or 50 km, or by census unit, such as inter-county or inter-state. For a power-law distributed measurement this introduces large inconsistencies: Figure 2b shows that defining “did not migrate” as birth/death within 10 km vs the average inter-county distance in Britain (56 km) or inter-commuting zone distance in the US (83 km) would give 20-30\% differences in apparent percent migration for the same dataset.

\iffiguresInText
    \begin{figure}[H]
        \includegraphics*[width=12cm]{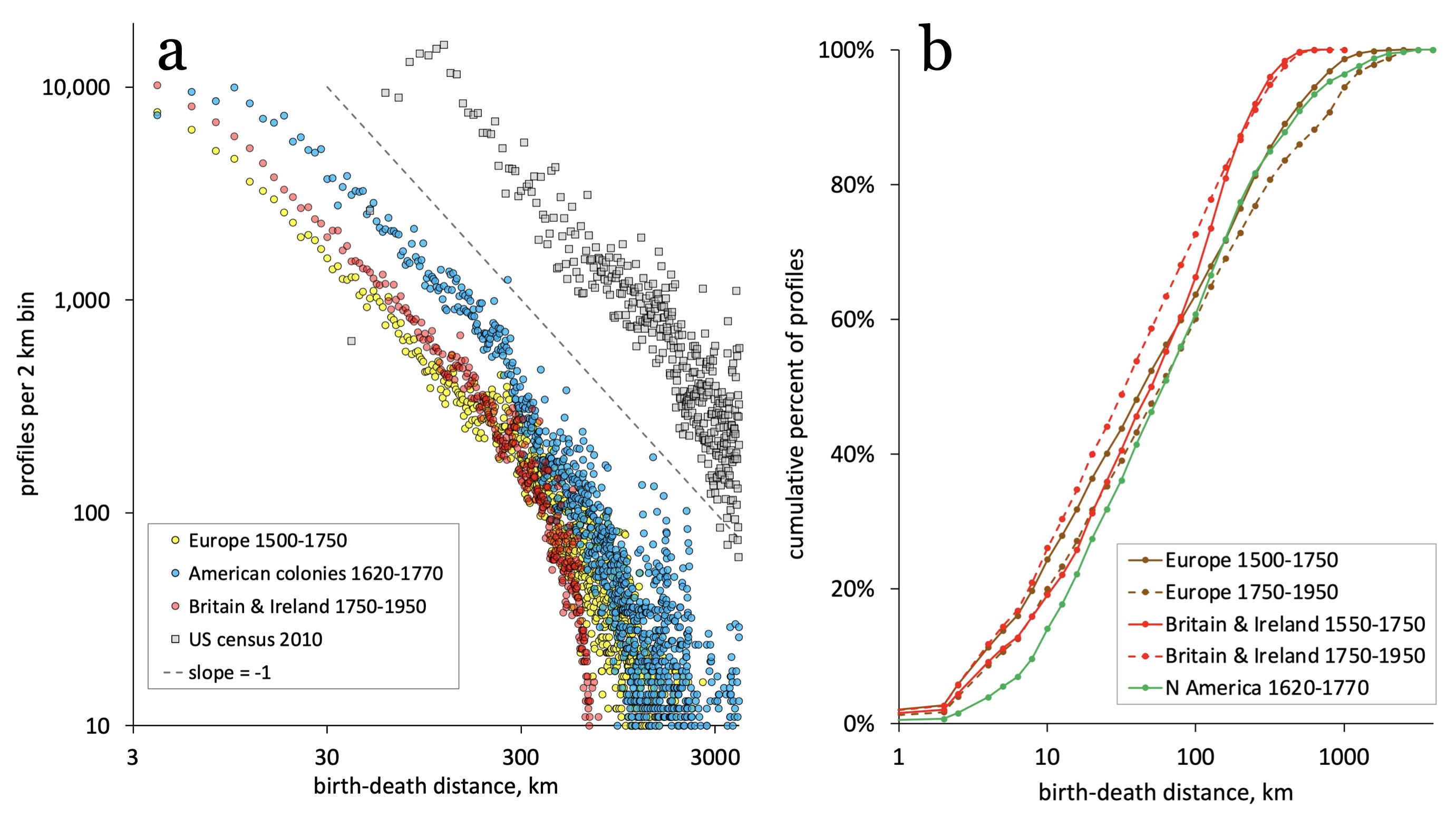}
        \centering
        \caption{a: circles: Distribution of birth-death distance, excluding trans-continental migration, for the indicated regional and temporal subsets of the Kaplanis dataset. squares: US census 2010 age 16-26 cohort distances.  The dashed line has slope = -1, for reference. b: Cumulative distributions of the Kaplanis data with additional time-slices, counting only those who moved at least 1 km.}
        \color{gray}\rule{10cm}{0.5pt}
    \end{figure}
\fi

The second consistent deviation from a 1/r distribution is the drop-off for all datasets at large distances, which reflects that the would-be migrant could go no further: for example London to Glasgow is 560 km, Spain to Germany 1600 km, Italy to Norway 2000 km, and Boston to St. Louis is 1600 km; these scales are reflected in the upper asymptotes of the datasets.

The third and most interesting deviation from the 1/r distribution is not shown in Figure 2; it begins in the US data after 1710, characterized by an order-of-magnitude increase in median migration distance with striking divergence from the power law, as shown in Figure 3 and discussed later.

\section*{The Inverse Distance (1/r) Distribution}

Because the data of Figure 2 precede the US frontier era and are taken from a wider geography, their pattern is the backdrop against which the American expansion can be contrasted.

Ancient Roman and early medieval city-sizes in Figure 1 puncture technology-based theories for city growth.  Figure 2 does the same for migration theories: whatever drives this pattern of human migration cannot depend on post-industrial communication or technology, since the distributions before and after 1750 are essentially identical. 

Temporal and geographic subsets of the Kaplanis data make it possible to exclude a variety of factors which might give rise to the observed distribution.  Firstly, the distribution is not dependent on population growth. From 1550-1750, European and British population growth was low, doubling every 360 years, while the American colonial population grew an order of magnitude faster, doubling every 24 years\footnote{Growth rates are readily calculated from population data, e.g. \url{https://en.wikipedia.org/wiki/Demographic_history_of_the_United_States}}.  Secondly, distributions of internal migration are separable from significant international migration, either as emigrants (from Europe) or immigrants (to North America).  Thirdly, the distributions are minimally dependent of gender: the differences between eras and regions greatly exceed the differences between female and male migration (SM2).

Power law distributions for travel have been observed before.  Zipf’s early rationale was that the optimization of inter-city transportation costs, assuming that such costs are proportional to distance (Zipf 1946), led to the observed 1/r dependence, though he measured frequent and reversible travel (train and bus passengers), not lifetime migration, and assumed a constant, steady-state population.  Others have observed power laws in travel rationalized by gravity models, radiation models, intervening opportunity models and more (Gonzalez, Hidalgo, \& Barabasi 2008, Levy 2010, Alessandretti, Aslak, \& Lehmann 2020, Simini et al 2012, Schläpfer et al 2021).  This literature finds exponents of either -1 or -2 and, like Zipf, is concerned with a modern era of commuting-scale reversible travel.  Most of the models are symmetric with travel flows proportional to the product of the origin and destination populations — but this is irrelevant to frontier scenarios that are very far from a steady-state with essentially zero population at the destination.
Given these contradictions, inapplicability to our medieval-to-modern timescales, empirical disagreement with slope -2 results, and possible irrelevance of commuting to migration behavior, we eschew these models as well as the premise that human behavior should have commonality with Newtonian masses or radiating particles. What matters here is only that the 1/r distribution of migration distance is a robust observation that can be parameterized, observed across time and geography, and examined for exceptions that illuminate history. 

\section*{The Late Colonial American Shift}

Figure 3 shows that from 1620 to 1710 the migration distribution follows the European pattern, but then from 1740 to 1800 — in only two generations — there are three dramatic shifts.  The vertical intercept shifts dramatically down as the number staying home drops from 40\% to 20\%.  At the same time, the maximum migration distance (the distance at which the distribution extrapolates to 100\% of the population) increases from ca. 200 km to 1000-2000 km.  Finally, the 1/r distribution collapses (the linearity of the semilog cumulative distribution function (CDF) vanishes) as many people move 50-100 km or more.  From 1860 onward, the distributions return partway to the colonial pattern, though total “leavers” remain in the 70-80\% range.

\iffiguresInText
    \begin{figure}[H]
        \includegraphics*[width=10cm]{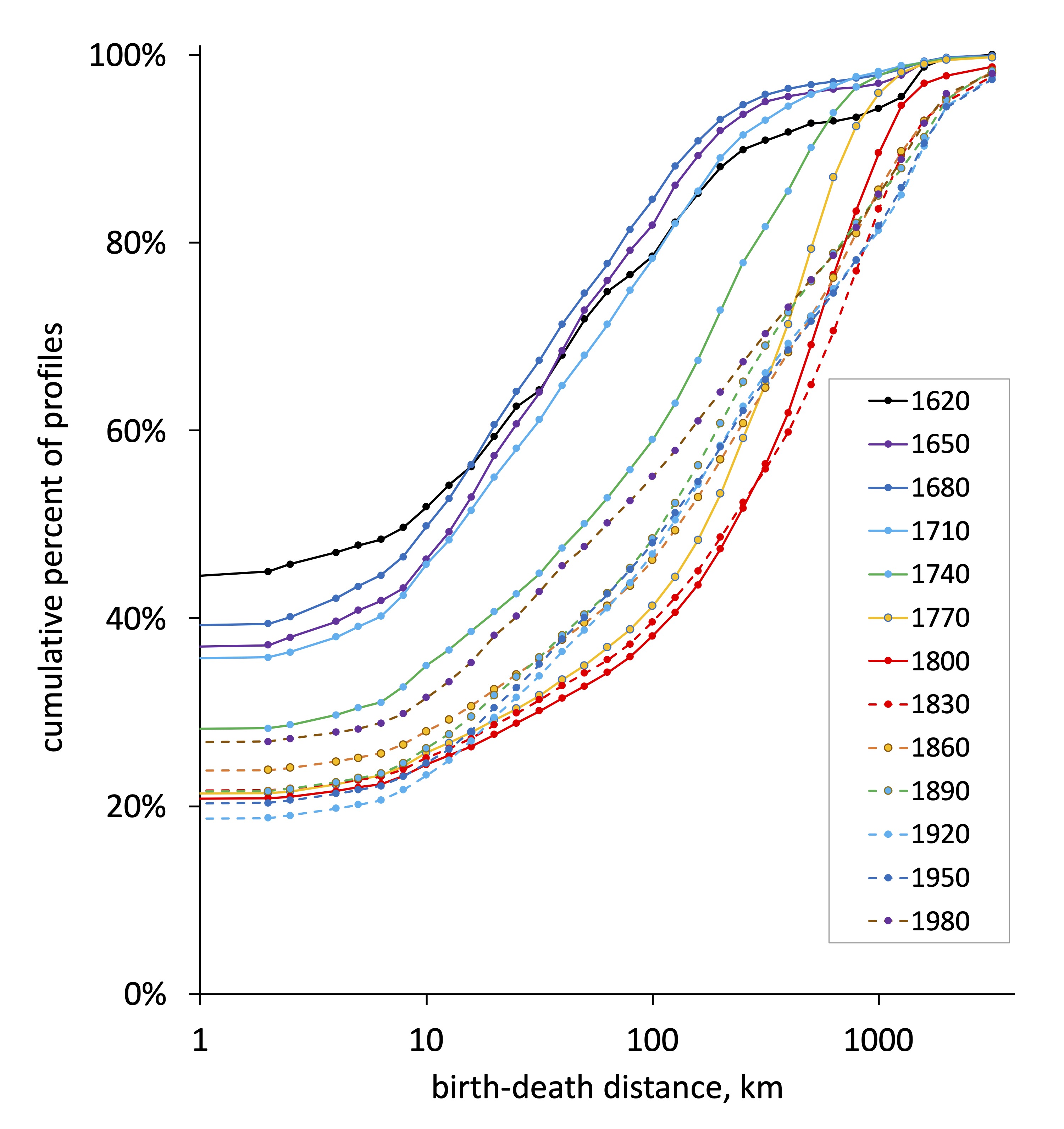}
        \centering
        \caption{Cumulative distributions of 2.83 million North American birth-death distances.  Unlike Figure 2, all profiles are counted, including those who did not move, so the percent of “stayers” appears as the vertical axis intercept.  Each curve is a 30-year (generational) subset of 20-year-olds; thus the 1620 data tallies those age 20 between 1620 and 1649. In all cases at least 100 profiles are counted in the smallest distance bin (2-2.5 km) from which counting proceeds in 10 bins per factor of ten.}
        \color{gray}\rule{10cm}{0.5pt}
    \end{figure}
\fi

\section*{Three Parameter Model}

The 1/r distribution in the European and early American data is so prevalent for so long that it is unlikely that its driving forces suddenly vanished around 1740; it is more likely that a new factor appeared.  A mathematical model can help to separate and quantify the transition.

Two parameters are obvious.  Britain and Ireland are geographically smaller than Europe or North America, and the British-Irish distributions plateau at a two- to three- fold lower distance than the others (Fig 2b); thus, one parameter should be a maximum distance, interpretable as a maximum physical or perceived range for migration. The second simple parameter is a fraction comprised of people who move at all (“leavers”) versus those who die very near their birthplace (“stayers”).  Even within the colonial era 1620-1710, the percent of leavers rose from 59\% in 1620 to 68\% in 1710 (Fig 3).  By fitting this value, we avoid arbitrary distance thresholds for stay-vs-leave.  The stay-vs-leave fraction is also the historic determinant of city-size distributions (Spencer 2024).

A third parameter is necessary to push the post-colonial US distributions to longer distances for a majority of people (Fig 3, after 1710).   The data suggest that this parameter should have units of distance, and since the distributions become short-tailed, that it should take the form of a fixed distance.  Decomposing migration cost into a linear combination of a fixed and distance-dependent component, integrating, and normalizing, gives a three-parameter model for migration distance (SM4):

\begin{equation}
      CDF = 1- f_{leave}(1 - ln( 1 + r/r_{fixed})/ln(1 + r_{max}/r_{fixed}))
\end{equation}

This is an empirical model, useful for decomposing and comparing migration across time and geography.  Fits are excellent (SM1, SM2, SM4, SM8) with rms error $\approx$3\%, suggesting that the principal determinants of migration distance are captured in three parameters.  A large and varied set of data slices shows no correlation between the three parameters (SM5).

Migration is usually defined and measured as the movement of people between administrative areas — counties, states, or countries — as recorded in censuses.  This gives rise to a long-standing problem: the diversity of those areas’ geographic and population sizes makes comparisons across time and space difficult or badly biased.  Viewing migration distances as a distribution and fitting that distribution to a model, e.g. Equation 1, can reduce and quantify that bias (SM6).

\section*{Overview of Migration History}

Figure 4 shows how these three parameters change over different continents and centuries, extending Kaplanis’ observations on median migration distance (Kaplanis 2018, figs S18, S19).  The percent leaving home (4a) falls from 65\% to 40-50\% from medieval to early 18th century Europe, then rises with rural-to-urban migration in the industrial revolution for Britain and Ireland and in the 19th century for continental Europe.  Americans’ $f_{leave}$ rises steadily from European values to 80\% by 1770, where it plateaus.

\iffiguresInText
    \begin{figure}[H]
        \includegraphics*[width=12cm]{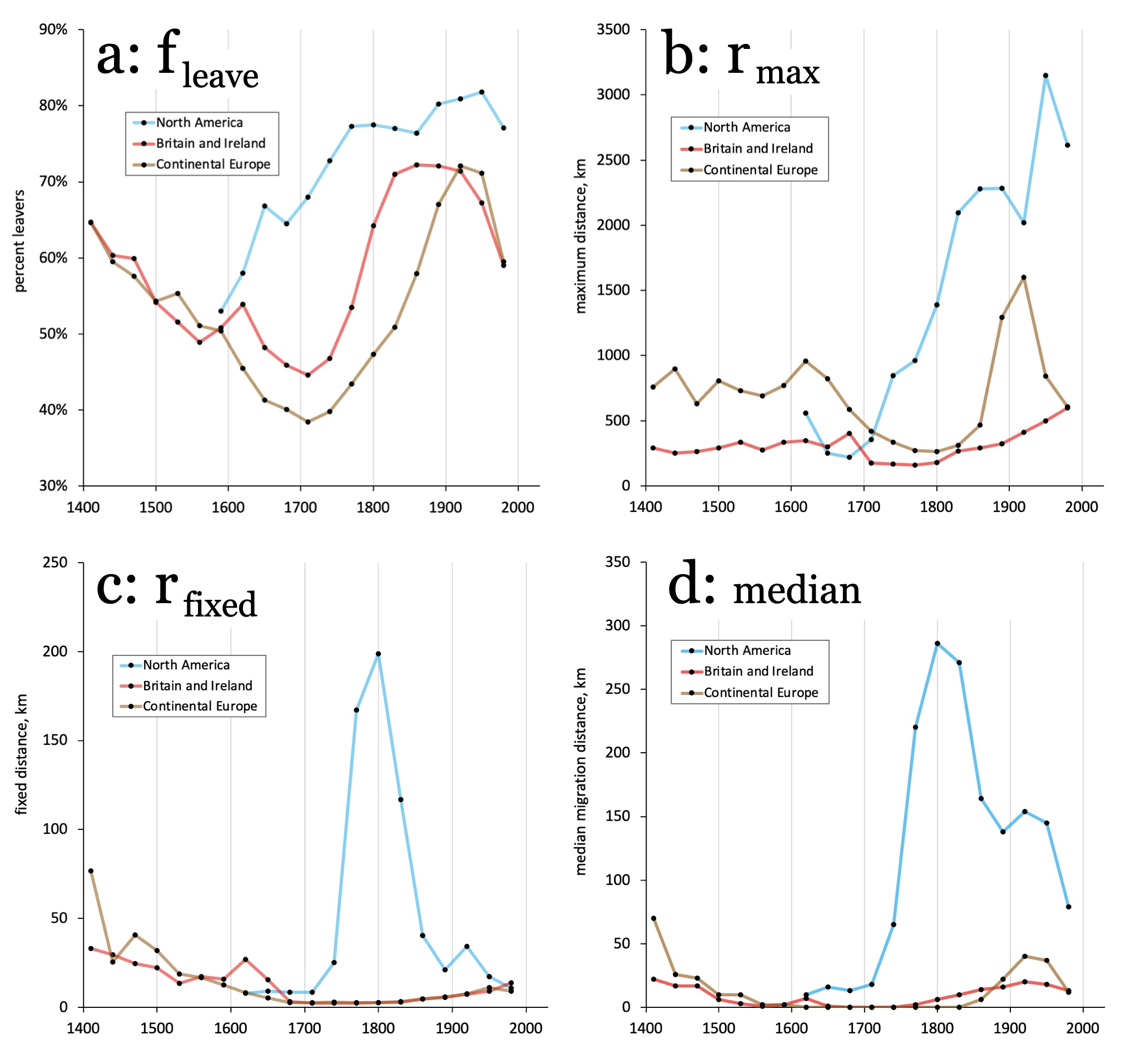}
        \centering
        \caption{Fitted values for $f_{leave}$, $r_{max}$, and $r_{fixed}$ of Equation 1, and observed median migration distance, by region and 30-year time slice.  Each point represents a minimum of 1500 profiles (median 12,900 for Britain \& Ireland, 149,700 for North America, and 63,100 for Europe).  Mean fit rms errors are 0.9\% for Britain \& Ireland, 1.3\% for North America, and 0.75\% for Europe.}
        \color{gray}\rule{10cm}{0.5pt}
    \end{figure}
\fi

The maximum extrapolated distance $r_{max}$, Fig 4b, is stable in Europe 1400-1650, when it curiously falls, only to rise 1860-1950.  This distance is essentially constant for the British Isles 1400-1980, reflecting the absence of barriers and finite size of these islands.  In North America the rapid rise of $r_{max}$ parallels the growing awareness of the opportunity (Lewis and Clark return in 1806) and is certainly cemented by the 1848–1855 Gold Rush that brought 300,000 people to California.

Values of third fitted parameter $r_{fixed}$ (Figure 4c) are low (0-25 km) and stable in Europe and the British Isles 1550-1980, and we suggest that this principally represents the distance resolution limit of the Kaplanis data (see SM7 for a census-based dataset with known and much greater resolution limit).  But the American $r_{fixed}$ displays a unique spike from 1730 to 1890; this cannot be due to a resolution limit since the data source is unchanged and the magnitude (200 km in 1800) is ten- to twenty- fold higher than the earlier baseline.  A fine-grained overlay of this era (Figure 5) emphasizes the independence of $r_{fixed}$ (red) and $r_{max}$ (blue) and shows that the observed median lifetime migration distance (black) can be decomposed into contributions from both.

\iffiguresInText
    \begin{figure}[H]
        \includegraphics*[width=12cm]{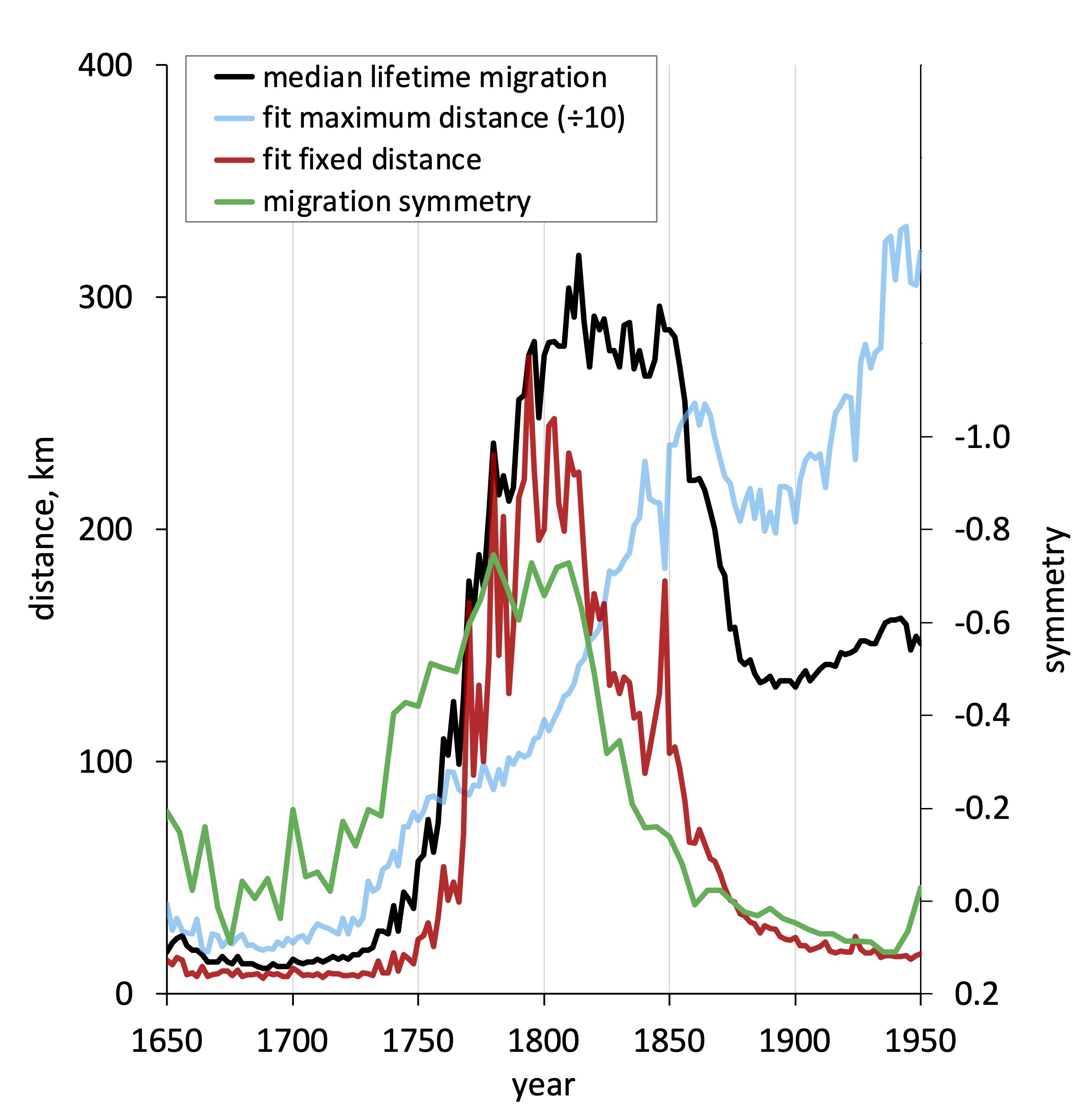}
        \centering
        \caption{Left axis: American distance parameters over time; 2.67 million profiles total, median 10,800 per 2-year slice.  The 1750-1890 spike of $r_{fixed}$ is shown in red, with $r_{max}$ in blue and measured median birth-death distance in black.  Right axis: Internal migration symmetry (green) as median of 1° geographic blocks with at least 40 births and deaths per 5 year period.}
        \color{gray}\rule{10cm}{0.5pt}
    \end{figure}
\fi

\section*{Decomposition by Geography}

Decomposition by time clarifies some of the determinants of migration distance, such as the leap in British and Irish “leavers” coincident with the Industrial Revolution noted above.  

One possible cause for the American spike of $r_{fixed}$ is that economic and technologic change made migration easier and less expensive than it had been, but this cannot account for the return of $r_{fixed}$ to its baseline by 1890. An alternative explanation is suggested by the geographic concentration of high $r_{fixed}$ in western New York state and east of the Appalachians from Virginia to South Carolina (SM8). These would-be migrants faced a significant barrier (Ontario or Lake Erie or the Appalachian mountains), but knew that attractive land lay beyond, and so incurred a “fixed cost” to cross that barrier.   They were also a decade older when they moved than earlier or later generations (SM3), suggesting that they needed time to acquire the resources for the longer jump.

Since our dataset records birth and death locations by latitude and longitude, not by county or state, we create and map data subsets in geographic blocks, typically 0.5°x0.5°\footnote{At 37°N a 0.5°x0.5° block has dimensions of 44x56 km = 2500 km$^2$. In the lower 48 states there are 3108 counties with mean area 2600 km$^2$. In effect these blocks are near constant-area counties.}.  The distributions of the profiles in such blocks with the highest and lowest values of $r_{fixed}$ are illustrative (SM8, lower chart).  The lower third, along the New England coast, have the same distribution as the colonies as a whole in 1710 — their behavior did not change and their migrations follow the 1/r distribution.  The upper third, in western New York and Virginia to South Carolina, have a short-tailed distribution with fitted $r_{fixed}$ of 548 ± 32 km (203,000 profiles, rms error 1.8\%) — a distance sufficient to jump the Appalachians or Lake Erie.

\section*{Vector Orientation}

With genealogic birth and death locations in latitude and longitude, the direction of each life-vector is readily calculated.  The dominant westward direction of American expansion is apparent (SM9), in good agreement with the directionality of the largest DNA clusters of Han et al (2017), figure 3.  The average early colonial vector is southwest (from Boston down the coast) but by 1800-1830 the arc is due west (270°) and remarkably narrow, then broadens back to colonial variance by 1920-1950. The narrow spread of the angular distribution (Figure 6, green) coincides with the spike in $r_{fixed}$ (red). Thus, the dominance of the western pattern of migration for that century plus (1800 to 1920) is neatly described.
  
\section*{Symmetry of Migration}

A symmetry metric is also readily found for any given region and time period by counting births and deaths, but only in different regions since we are concerned with migration and not population growth:
	$S = (births - deaths)/(births + deaths)$

Migration is at a steady-state when $S$ equals zero; $1 \geq S > 0$ indicates net efflux (more people leave than arrive) and $0 > S \geq -1$ indicates net influx (SM10).  The absolute value of $S$ measures net migration, and the mean of $\lvert S \rvert$ for a large area’s composite blocks (e.g. counties or geographic blocks) is a measure of whether it is near or far from a steady state.  Figure 6 shows Great Britain, Ireland, and Europe in a near-steady state for the past 600 years.  Certainly, there is asymmetric migration in specific times and places, for example rural Britain to industrial cities after 1750, but overall internal migration is close to reciprocal.  Again, North America is very different with much higher “churn” from the colonial era to the present; it is a system far from a steady-state.  Decomposing by geography (maps SM10, SM11) shows that the unique peak of mean absolute symmetry 1710-1890 is due to migration into essentially empty areas (as far as the settlers were concerned) that could not be reciprocal. The sign of $\Sigma S$ shows that 1710-1890 migration was dispersive (frontier) and crossed to concentrative about 1900 (Figure 6, yellow dots; full distributions SM10) as internal American migration became dominated by urbanization and moves to warmer climes (SM10).  The recent baseline (1920-1980, $S \approx$ 0.25) remains 2-3 fold above the European norm; America is a dynamic system relatively far from equilibrium.

\iffiguresInText
    \begin{figure}[H]
        \includegraphics*[width=12cm]{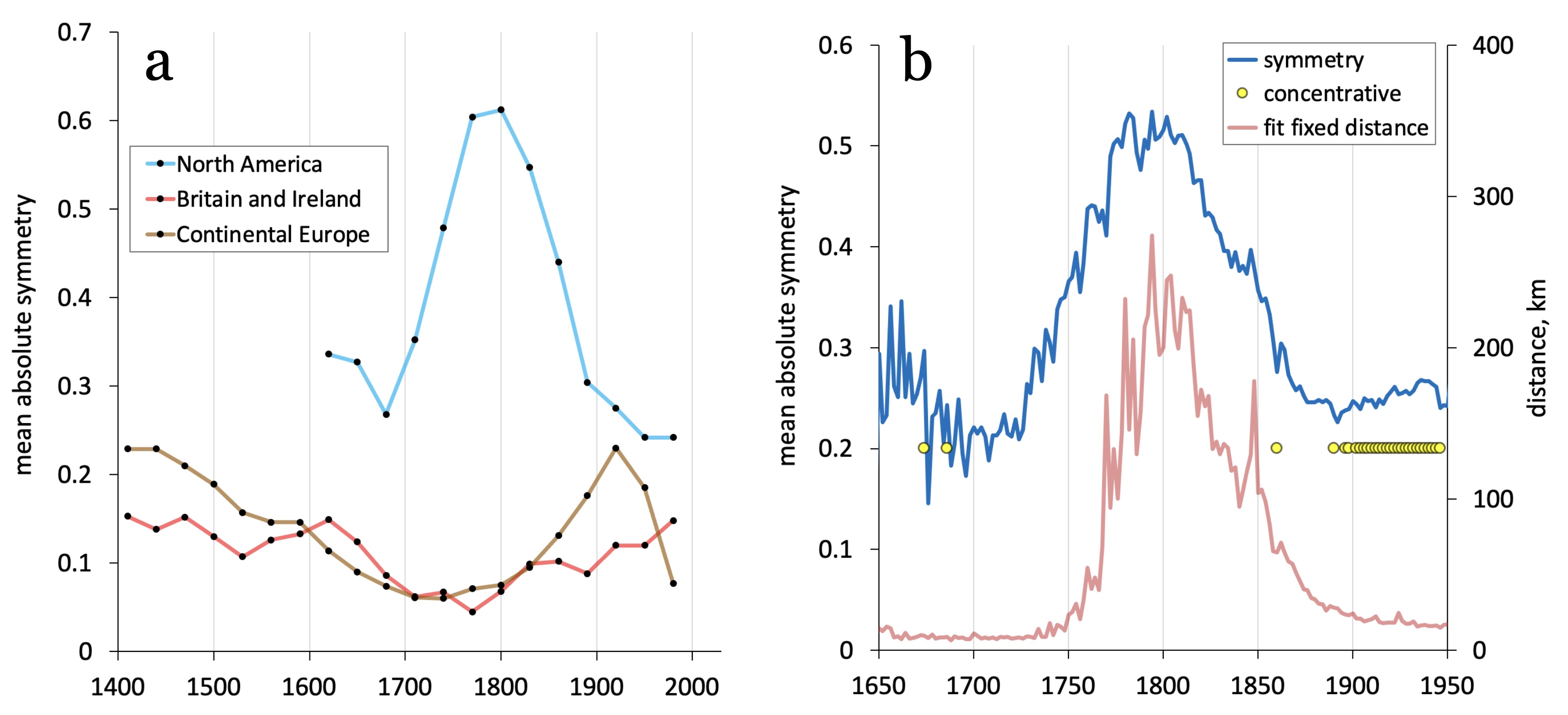}
        \centering
        \caption{Migration symmetry by region and date.  Values are mean absolute symmetry = $\Sigma\rvert$S
    $\rvert/n$ over all 1x1° blocks in (a) the indicated region in 30 year time slices and (b) North America in 2 year slices.  Panel (b) repeats the data for $r_{fixed}$ of Figure 5 to show their temporal overlap.  Yellow dots in (b) indicate time slices for which $\Sigma S$ is positive, i.e. migration concentrates the population; otherwise $\Sigma S$ is negative or dispersive.}
        \color{gray}\rule{10cm}{0.5pt}
    \end{figure}
\fi

\section*{Discussion}

That migration distance follows a robust inverse-distance power law is not new, having been documented from Zipf (1946) to Lehmann (2024).  Our contribution is the observation that this pattern has persisted at least from the 16th to the 21st centuries and across two continents (Fig 2) which allows us to say what does \emph{not} give rise to this pattern. It does not depend on post-industrial communication or transportation. It does not depend, as required by gravity models, on a reciprocal pull from distant population centers. It does not depend on reversible exploration.  In addition, it does not depend on population stasis or growth, nor on the gender of the travelers.  The origin of the 1/r dependence is uncertain but must be pre-industrial and cross-cultural; it may arise in our innate logarithmic perception of distance and cost (Boucherie, Maier, \& Lehmann 2024, Dehaene 2003) and even be related to similar distributions of other species’ foraging behavior (Gonzalez, Hidalgo, \& Barabasi 2008).

Taking an underlying power law with slope -1 as a given, migration distance can then be decomposed into a model with only three fitted parameters; fitting a wide range of temporal and geographic data subsets shows that the model is widely applicable (root mean square error 2-4\%) and the parameters are uncorrelated.  The distribution of migration distance retains its 1/r character over the entire observed range of two of the parameters, namely from 40\% to 70\% of young adults leaving their birthplace ($f_{leave}$), and a maximum distance $r_{max}$ ranging from hundreds to thousands of kilometers. The distribution becomes short-tailed (the power law breaks down; the distance range is narrow) when the fixed-cost term $r_{fixed}$ is well above the intrinsic resolution of the data (ca. 10-20 km, SM6) which occurs only during the American expansion from 1770 to 1860 (Figure 6) and only in specific locations (SM8).

\section*{Mapping the Frontier}

The American frontier is usually mapped and delimited by population density from census data, typically as the line dividing areas of 2 or more people per square mile from areas with fewer (Gannett 1903; Otterstrom and Earle 2002; Bazzi, Fiszbein \& Gebresilasse 2020).  Yet population density is a metric of convenience: deserts and bodies of water have very low population densities, but their boundaries would not be considered frontiers because they are stable.  A frontier is defined by the movement of people, and genealogic data offer alternatives which directly reflect that fundamental.

A frontier arguably ends when people “settle down” and stop moving, for which an appropriate measure is the fraction of those who do not leave their birthplace, i.e. $1 - f_{leave}$ ;  a value of 10\% staying roughly corresponds to the classic census frontier line (SM11).

Migration symmetry $S$ may also be used to delimit the frontier.  Frontier expansion is just that — expansion, namely the movement from higher to lower density, therefore requiring $\Sigma S > 0$ for regions behind the frontier and $\Sigma S < 0$ ahead of it.  The $S = 0$ line is quite a good frontier marker (SM11).  Mapping symmetry has the additional advantage of showing where people came from and where they went to.

These two metrics — the fraction staying and migration symmetry — are intrinsically about people moving.  They are independent of baseline population growth.  Symmetry reverts to a steady-state value after the frontier passes by.  They are available from genealogies starting in medieval Europe and 1620 in North America — six generations before we can visualize the frontier with census data.  Indeed, it is noteworthy that these measures do not depend on population density to demarcate the frontier line, as other studies have.

\section*{Metrics and Properties that Distinguish the Frontier}

Our distance, directionality, and symmetry parameters have stable and comparable values for early colonial North America and Europe 1400-1950. We use the term \textit{frontier}\footnote{We acknowledge that what can be called frontier migration from one side can be called an invasion from the other. Our story is necessarily one-sided due of the availability of genealogic data. No bias or judgment is intended.} to denote places and times when their values are markedly different.  Table 1 summaries the properties of what we call this frontier pattern vs. its steady-state alternative and provides a framework for the following discussion.

\iffiguresInText
    \begin{table}[h]
    \centering
    \begin{tabular}{|c|c|}
    \hline
    \multicolumn{1}{|c|}{{\color[HTML]{0080D8} steady state}} & \multicolumn{1}{|c|}{{\color[HTML]{0080D8} frontier}}  \\ \hline
    
    \multicolumn{2}{|c|}{\color[HTML]{0080D8} migration distance CDF fitted parameters}  \\
    $f_{leave}$ 40-70\%       &    $f_{leave}$ 70-90\%         \\
    $r_{max}$ 200-1500 km     &    $r_{max}$ 1000-3000 km      \\
    $r_{fixed}$ 0-40 km       &     $r_{fixed}$ 50-200 km      \\
    
    \multicolumn{2}{|c|}{\color[HTML]{0080D8}additional measurables}  \\
    median distance 10-20 km  &    median distance 150-200 km  \\
    no common direction       &    narrow common direction     \\
    symmetric                 &    asymmetric                  \\
    \multicolumn{2}{|c|}{\color[HTML]{0080D8}inferences}         \\
    median travel time $\sim$days &     weeks/months       \\  
    may be reversible         &     irreversible       \\
    destination known         &     destination unknown       \\
    family ties possible      &     family ties broken       \\
    sustainable               &     self-extinguishing       \\
    optimizing                &     opportunistic       \\
    gravity models may apply  &     gravity premise invalid       \\
    universal, medieval to modern &  rare but lasting effect       \\
    
    \multicolumn{2}{|c|}{\color[HTML]{0080D8}common properties}  \\
    \multicolumn{2}{|c|}{peak migration $\sim$age 20} \\
    \multicolumn{2}{|c|}{independent of population growth, technology, or gender} \\
    \hline
    \end{tabular}
    \caption{Properties that Distinguish Steady State and Frontier Migration}
    \end{table}
\fi

\subsection*{The decision to leave}

The conventional wisdom is that American young adults have a greater tendency to leave home than Europeans, though Long \& Boertlein (1976) assert that “statistics which permit accurate comparisons among countries in this respect are of very recent origin.”  We disagree: genealogies yield quantitative measures that span centuries. Colonial American $f_{leave}$ begins its steep rise from $\sim$50\% to 80\% almost immediately (Fig 4a).  One contributing factor could have been a desire to land-grab before the wilderness was all claimed and parceled.  Additionally, the premise of the 1/r distribution is that there is some sort of cost, proportional to distance, associated with leaving home.  If the emotional portion of that cost — leaving behind family and the familiar — had already been paid by trans-Atlantic emigration, and/or if settlement was so fast and sparse that cohesive towns had not developed, then there would be weaker ties for colonial young adults, resulting in both a lower barrier to leave home and a reduced cost associated with long distance.  Fischer (1989) notes the regional and cultural differences in colonial America, namely that those in the Southern Highlands were far more likely to move (persistence 25-40\%) than New Englanders (75-96\%) and had a relaxed approach: “When I get ready to move, I just shut the door, call the dogs, and start.” (Fischer p 759).  Our data show that from 1740 to 1800 that high propensity to leave was indeed concentrated in Virginia and the Carolinas, but after 1830, 80-90\% of young adults leaving home was widespread\footnote{See \url{http://scaledinnovation.com/gg/migration/migration.html?type=fleave&loop}}.

\subsection*{Independent decisions}

Since there is clearly a strong distance dependence to migration, could the stay-vs-leave decision be an artifact of the 1/r distribution?  That is, are “stayers” simply those whose migration distance falls on the 1/r distribution but is too small to have been measured?  The 2010 census study is suggestive since the high percentage of stay-at-homes (70\%) drops significantly (to 40-50\%) if the distance-unit is extrapolated from a commuting zone ($\approx$80 km) to a town-scale distance ($\approx$10 km) (SM7).
However, we believe the two key decisions (stay-vs-leave, how-far) to be independent.  The modeled parameters are uncorrelated (SM5).  Their timing is uncorrelated:  Britain and Ireland 1740-1830 and North America 1620-1710 see a large rise in leavers but no significant change in distance (Fig 4).  We suggest that the stay-vs-leave decision is significantly driven by local factors (\textit{How many siblings do I have? Can I inherit the farm or business?}) while the how-far decision is driven by distance-related factors (\textit{Where are the risks and opportunities? Can I rely on anyone there? How will I travel?}).

\subsection*{Maximum distance}

Internal migration has been constrained by geography in Britain and Ireland, and by political boundaries in Europe, for most of recorded history. But in North America, a continuing series of explorations and events (the Great Lakes in the 17th century, Boone’s Wilderness Road 1775, Lewis and Clark 1804-1806, Sutter’s Mill 1848) grew the maximum migration distance from British scale ($\sim$200 km) to continental scale ($\sim$2500 km) in four generations (Figure 4b).

\subsection*{Fixed distance}

The geographic locations of the 1770-1860 spike in $r_{fixed}$ suggest that this term records a jump over a barrier, essentially a fixed-cost that precludes short migration distances and therefore disrupts the 1/r pattern.  There is another large physical barrier encoded in the genealogies, namely the unavoidable 5000 km transatlantic voyage of European emigrants to North America. We do not include these immigrant profiles in our analysis for several reasons: Firstly, trans-national migration is well studied so we focus on internal migration.  Secondly, these profiles disrupt distance distributions to the point that they are no longer monotonic, and with the additional term(s) or rules required for their modeling there would be no net gain of understanding.  These profiles may also be prone to selection bias, since the Kaplanis data draws from Geni.com, an American crowd-sourced genealogy site likely to select for immigrant vs European local migrant profiles.  That said, the life-death distances of those born in Ireland 1790-1850, and therefore alive during the potato famine, show fixed-distance terms consistent with crossing water barriers to Britain, to North America, and to Australia.

A large fixed-cost term would not seem essential to a frontier: it is easy to imagine a large-scale, asymmetric, irreversible, directional movement of people without a major geographic barrier.  And yet the best such examples may be prehistoric (see below); perhaps we humans pushed across all of the easy (walkable) routes millennia ago, giving Eurasia time to settle into a steady-state until technology (ocean-crossing ships) broke down those barriers to unleash another frontier in the Americas.

\subsection*{Symmetry and directionality}

Gravity and related models assume a migration steady-state and encode symmetry in their numerators with migration between A and B proportional to (population of A)$\cdot$(population of B) (Simini et al 2021 equation 1;  Barbosa et al 2018 equation 4).  Strictly speaking, mean absolute symmetry $S$ should be zero under a gravity model, but we can take six centuries of relatively stable British-Irish and European data with $0.5 < S < 0.2$ to represent a realistic steady-state for internal migration (Figure 6a).  Americans are not just more mobile (Figure 4), they are farther from equilibrium with $0.25 < S < 0.6$, a substantial violation of the gravity premise.  The spike of $S$ begins $\sim$1740, thirty years before that of $r_{fixed}$ (Figure 6, right); this early phase goes up northern rivers and into central Virginia (SM10, SM11), relatively easy geographically and so consistent with interpretation of the $r_{fixed}$ spike caused by must-cross barriers.

Horace Greeley’s admonition was well founded: the spikes of $r_{fixed}$ and $S$ coincide precisely with a very narrow due-west distribution of migration direction (SM9) that reflects a similar directionality toward opportunity and geographic paths.  

\subsection*{Implications}

We refer to the historically dominant (non-frontier) pattern of migration as steady-state: its power law distribution reflects a majority of short moves and a few long moves, as evinced by median migration distances under 50 km (Figure 4d), a distance that can be covered on foot in a few days.   Short distances mean that migration can be reversible and need not sunder family and cultural ties.  It can be exploratory and then to a known destination; in the last two centuries it is most often city-to-city.  We agree with the large literature that such migration is driven by personal optimization.  As an individual behavior it has no coordinated common direction.  At a true steady-state such migration must be symmetric, as gravity models demand, and as such it is indefinitely sustainable.

Though the American migration distance metrics are very different from the European baselines (Figure 4), by themselves these metrics do not require a frontier.  But their coincidence in time and place with strong asymmetry and directionality — all derived from the same genealogies — assure that these descriptors do, in combination, represent a consistent frontier pattern.   Importantly, most of these metrics revert to early colonial or European levels by 1860-1890, signaling the end of the frontier era.

The frontier metrics have strong implications.  Median migration distances of 150 to 300 km (Figure 4d, 5) require significant effort and time, making it functionally irreversible (before modern transportation) and breaking family ties. Destinations might be known in general (broad woodland or prairies) but not specifically.  There was no one (excluding the indigenous people) at the far end, precluding symmetry.  Very narrow directionality (SM9) implies coordination; people were communicating about opportunities and options and not acting independently (i.e. unlike later urbanization which is largely isotropic).  The frontier pattern is not sustainable but self-extinguishing.

By our data-based definition, American frontier migration ends with the return-to-baseline of $r_{fixed}$, directionality, and symmetry, all during 1860-1890.  But two metrics, $r_{max}$ and $f_{leave}$ , remain high despite frontier closure, the first because knowledge cannot be unlearned (California is always an option for a New Yorker and vice-versa) and the second because coincidental appearance of inexpensive long-distance communication and transportation keeps young adults aware of opportunities while decreasing cultural or family disruption.

\section*{Preceding Frontier Events}

Every frontier migration is an invasion but not every invasion is a frontier migration. Napoleon’s westward path across Europe and Russia makes the distinction: Like frontier migration it was rapid, highly directional, asymmetric, and of limited duration, but it was not gender-balanced and did not result in significant settlement by the newcomers.  It did not have a major irreversible effect on the invaded people as measured by their genealogies or DNA.  

Our frontier migration pattern (Table 1) has more in common with mass migration to avoid war or famine: a long jump to the unknown that is irreversible, unidirectional, and of relatively short duration.  However, two elements distinguish such avoidance migration from frontier migration. Firstly motivation, since one is push, movement away from something, while the other is pull, movement toward something. Secondly, avoidance migration is not generally into an empty or underdeveloped area, but rather into an accommodating civilization.  

But more complete parallels to frontier migration may be found in prehistory.  DNA analysis convincingly shows that two ancient transitions in Europe — Mesolithic to Neolithic, then Neolithic to Bronze Age — were \textit{demic}, driven by the movement of people (Haak et al. 2015, Olalde et al. 2018, Brace et al. 2019, Chikhi et al. 2002).  These were irreversible, directional, frontier migrations of people into regions of low population and with the advantage of new technologies: domesticated plants and animals from the Near East and Anatolia, and then bronze tools, language, and horses from the Steppes.  The newcomers substantially (Neolithic) and essentially completely (Bronze Age) replaced the locals.  The parallels to the frontier era of North America are striking, including reversion to steady-state behavior once the frontier reached the western ocean boundary (Antonio et al 2024).

\section*{Questions Raised and Next Steps}

Millions of genealogic profiles show that the distribution of lifetime migration distance persists across centuries.  That the pattern is pre-industrial and cross-cultural eliminates a host of modern hypotheses for its root cause.  Genealogies also provide migration distance, directionality, and symmetry, which form a set of quantitative descriptors that characterize frontier migration.  Such metrics cannot be expected to provide every answer, but can by mapped against known history and can raise new questions.   For example:
\begin{itemize}
\item The swift rise of $f_{leave}$ from 45\% to 70\% in Britain and Ireland 1700-1830 almost certainly reflects migration for new urban, industrial employment, yet it is not accompanied by any change in $r_{max}$ or $r_{fixed}$ (Figure 4).  A similar rise of $f_{leave}$ from 40\% to 70\% occurs in Europe 1700-1900, but with (1860-1900) a 5-fold rise in $r_{max}$ .  Is this also about urban employment, perhaps to less dispersed locations, or does it reflect political changes (war)?
\item The location and timing of the 80-90\% leavers and $>$100 km $r_{fixed}$ long-jumpers we see in 1740-1800 Virginia suggest that this is the Scotch-Irish folkway (Fischer 1989) which begat the stereotypic Appalachian culture.  Yet a majority did not stop in the mountains but traveled farther and continued westward (Figure 7a,b).  What are the details of these different paths from a common origin? 
\item From 1860-1920 there is a strong arc from Boston, across northern Ohio, into Minnesota, of very low $r_{fixed}$ , i.e. nearly perfect 1/r distributions (Figure 7c). Most leave home (70-90\%) yet travel only 10-30 km.  Why does this narrow upper Midwest arc have the same pattern as coastal New England, and differ from that south of mid-Ohio?
\end{itemize}

\iffiguresInText
    \begin{figure}[H]
        \includegraphics*[width=12cm]{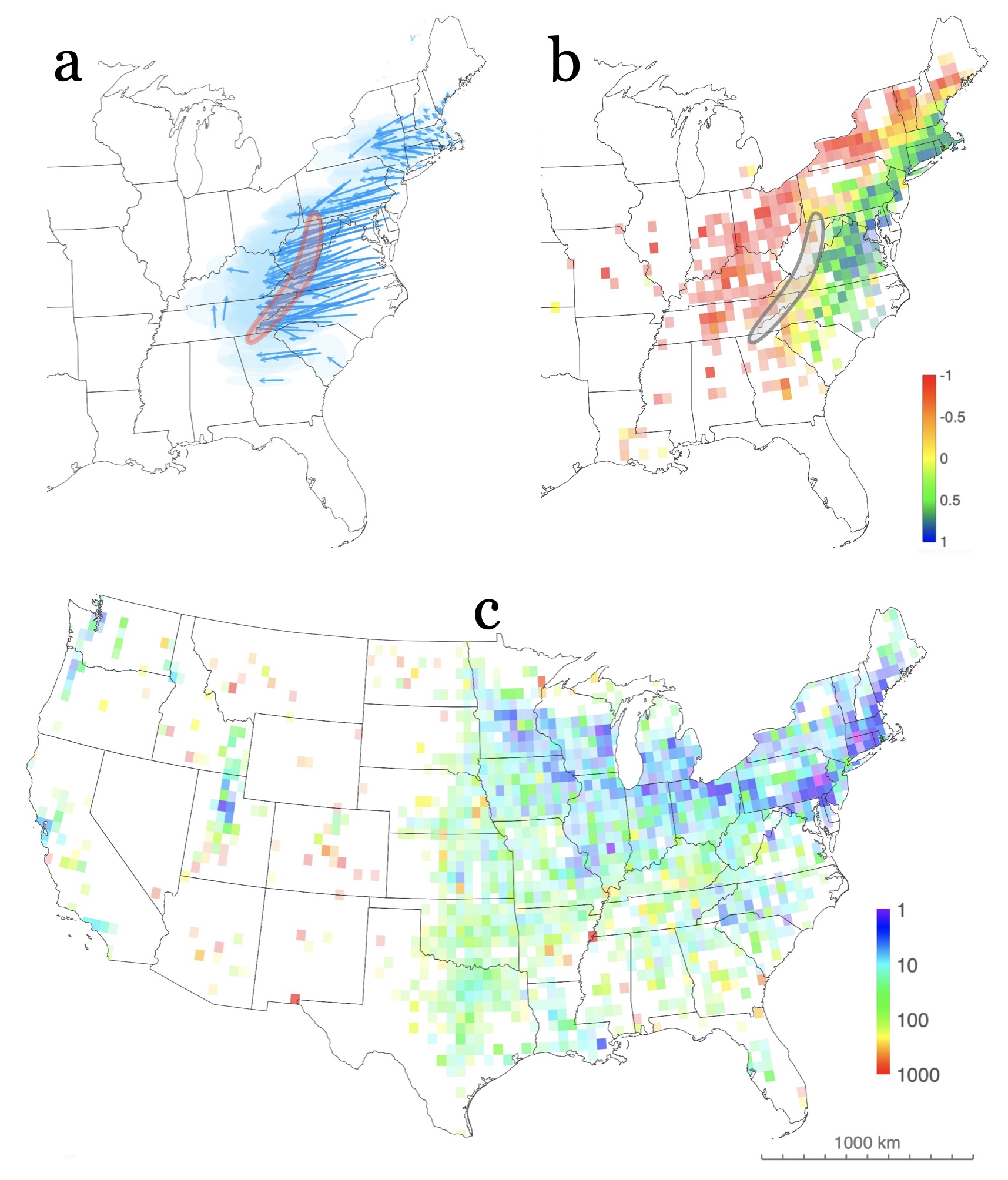}
        \centering
        \caption{Maps are computed for 0.5°x0.5° blocks with at least 100 profiles. (a) Median birth-death vectors, tally year 1770; light blue ellipses are ±1 standard deviation of the death locations of those born in the block at the arrow’s origin. 
    Note that a majority of the Virginia-origin vectors cross the Appalachian Mountains (oval shape).  (b) Values of symmetry $S$ for 1770 also showing that a majority of this generation dies well west of the mountains (red squares, $S \approx -1$, thus the earliest arrivals) (c) Values of $r_{fixed}$ for 1920. Note the blue arc from the New England coast, across Pennsylvania, Ohio, and Michigan, up to Minnesota.  Most of these people leave home ($f_{leave}$ = 0.7-0.9) but migrate short distances (median $<$ 30 km) with no fixed-cost penalty ($r_{fixed}$ $<$ 10 km).  These and more maps may be found at http://scaledinnovation.com/gg/migration/migration.html .}
        \color{gray}\rule{10cm}{0.5pt}
    \end{figure}
\fi

\break

The power of massive crowd-sourced genealogies can be harnessed to address these and other questions.  In particular, the anonymous profiles of the Kaplanis dataset can be matched generally by time and place, and in some cases exactly by tree-matching, to detailed profiles in FamilySearch and WikiTree.  Those additional names, life-stories, and historic connections can bring this history to life.

\section*{Data Sources and Methods}

Our sources are the curated and anonymized familinx dataset derived from crowdsourced genealogies (Kaplanis et al. 2018) and the US census study of Sprung-Keyser et al. (2022).  The genealogic data takes the form of birth-death distances in km, which for parameter estimation are tallied into logarithmic bins, summed, and normalized to give a CDF.  Fitting to Equation 1 is by Nelder-Mead optimization with values above 90\% excluded to avoid end-of-migration-range truncation. All fits have rms error < 4\%.  On PDF and CDF plots the datapoints have no error bars since they are simply counts of genealogy profiles.  
In order to compare time-slice analysis and mapping to traditional census-based analysis, genealogic profiles should counted in a way that reasonably corresponds to a census.   For a given “tally year” and geographic block, we count those born in that block if age 20 or younger at that tally year, and those who died in that block if age over 20. 

Source code, maps, and links to primary data sources may be found at \url{http://scaledinnovation.com/gg/migration/migration.html}.

\pagebreak
\section*{References}
\begin{hangparas}{5mm}{1}

Adams, J. W., Kasakoff, A. B. \& Kok, J. (2002). Migration over the life course in XIXth century Netherlands and the American north: A comparative analysis based on genealogies and population registers. Annales de Démographie Historique, 2, 5-27.	http://dx.doi.org/10.3917/adh.104.0005

Alessandretti, L., Aslak, U. and Lehmann, S. (2020) The scales of human mobility. Nature, 587(7834), pp.402-407.

Antonio, M. L., Weiß, C. L., Gao, Z., Sawyer, S., Oberreiter, V., Moots, H. M., ... \& Pritchard, J. K. (2024). Stable population structure in Europe since the Iron Age, despite high mobility. Elife, 13, e79714.

Barbosa, H., Barthelemy, M., Ghoshal, G., James, C. R., Lenormand, M., Louail, T., ... \& Tomasini, M. (2018). Human mobility: Models and applications. Physics Reports, 734, 1-74.

Batty, M. (2013). A theory of city size. Science, 340(6139), 1418-1419.

Bazzi, S., Fiszbein, M., \& Gebresilasse, M. (2020). Frontier culture: The roots and persistence of “rugged individualism” in the United States. Econometrica, 88(6), 2329-2368.

Blazquez, D., \& Domenech, J. (2018). Big Data sources and methods for social and economic analyses. Technological Forecasting and Social Change, 130, 99-113.

Boucherie, L., Maier, B.F. and Lehmann, S. (2024) Decomposing geographical and universal aspects of human mobility. arXiv preprint arXiv:2405.08746.

Brace, S., Diekmann, Y., Booth, T. J., van Dorp, L., Faltyskova, Z., Rohland, N., ... \& Barnes, I. (2019). Ancient genomes indicate population replacement in Early Neolithic Britain. Nature ecology \& evolution, 3(5), 765-771.

Chikhi, L., Nichols, R. A., Barbujani, G., \& Beaumont, M. A. (2002). Y genetic data support the Neolithic demic diffusion model. Proceedings of the National Academy of Sciences, 99(17), 11008-11013

Chong, M., Alburez-Gutierrez, D., Del Fava, E., Alexander, M. \& Zagheni, E. (2022) Identifying and correcting bias in big crowd-sourced online genealogies. Max Planck Institute for Demographic Research.

Colasurdo, A., \& Omenti, R. (2024). Using Online Genealogical Data for Demographic Research: An Empirical Examination of the FamiLinx Database (No. 62yxm). Center for Open Science.

Dehaene, S. (2003) The neural basis of the Weber–Fechner law: a logarithmic mental number line. Trends in cognitive sciences, 7(4), pp.145-147.

Deville, P., Linard, C., Martin, S., Gilbert, M., Stevens, F. R., Gaughan, A. E., ... \& Tatem, A. J. (2014). Dynamic population mapping using mobile phone data. Proceedings of the National Academy of Sciences, 111(45), 15888-15893.

Fischer, D. H. (1989). Albion's seed: Four British folkways in America (Vol. 1). America: A Cultural History, pp 759, 814.

Gabaix, X., \& Ioannides, Y. M. (2004). The evolution of city size distributions. In Handbook of regional and urban economics (Vol. 4, pp. 2341-2378). Elsevier.

Gannett, H. (Ed.). (1903). Twelfth Census of the United States, Taken in the Year 1900, William R. Merriam, Director: Statistical Atlas. United States Census Office.

Gonzalez, M.C., Hidalgo, C.A. \& Barabasi, A.L. (2008) Understanding individual human mobility patterns. Nature 453(7196), pp.779-782.

Haak, W., Lazaridis, I., Patterson, N., Rohland, N., Mallick, S., Llamas, B., ... \& Reich, D. (2015). Massive migration from the steppe was a source for Indo-European languages in Europe. Nature, 522(7555), 207-211.

Han, E., Carbonetto, P., Curtis, R. E., Wang, Y., Granka, J. M., Byrnes, J., ... \& Ball, C. A. (2017). Clustering of 770,000 genomes reveals post-colonial population structure of North America. Nature communications, 8(1), 14238.

Kaplanis, J., Gordon, A., Shor, T., Weissbrod, O., Geiger, D., Wahl, M., ... \& Erlich, Y. (2018). Quantitative analysis of population-scale family trees with millions of relatives. Science, 360(6385), 171-175.   Mirror data site \url{https://osf.io/fd25c/}

Koylu, C., \& Kasakoff, A. (2022). Measuring and mapping long-term changes in migration flows using population-scale family tree data. Cartography and Geographic Information Science, 49(2), 154-170.

Levy, M. (2010) Scale-free human migration and the geography of social networks. Physica A: Statistical Mechanics and its Applications, 389(21), 4913-4917.

Long, L. H., \& Boertlein, C. G. (1976). The geographical mobility of Americans: an international comparison (No. 64). US Department of Commerce, Bureau of the Census.

Olalde, I., Brace, S., Allentoft, M. E., Armit, I., Kristiansen, K., Booth, T., ... \& Reich, D. (2018). The Beaker phenomenon and the genomic transformation of northwest Europe. Nature, 555(7695), 190-196.

Otterstrom, S. M., \& Earle, C. (2002). The Settlement of the United States from 1790 to 1990: Divergent rates of Growth and the End of the Frontier. Journal of Interdisciplinary History, 33(1), 59-85.

Otterstrom, S. M., \& Bunker, B. E. (2013). Genealogy, migration, and the intertwined geographies of personal pasts. Annals of the Association of American Geographers, 103(3), 544-569.

Ravenstein, E. G. (1885). The laws of migration. Journal of the Statistical Society of London, 48, No. 2 (Jun., 1885), pp. 167-235

Rogers, A., \& Castro, L. J. (1981). Model migration schedules. Research Report RR-81-30. Laxenburg: International Institute for Applied Systems Analysis.

Schläpfer, M., Dong, L., O’Keeffe, K., Santi, P., Szell, M., Salat, H., ... \& West, G. B. (2021). The universal visitation law of human mobility. Nature, 593(7860), 522-527.

Simini, F., González, M. C., Maritan, A., \& Barabási, A. L. (2012). A universal model for mobility and migration patterns. Nature, 484(7392), 96-100.

Spencer, R. (2024). City size distributions are driven by each generation's stay-vs-leave decision, \url{https://arxiv.org/abs/2405.02129} [physics.soc-ph]

Sprung-Keyser, B., Hendren, N., \& Porter, S. (2022). The radius of economic opportunity: Evidence from migration and local labor markets. US Census Bureau, Center for Economic Studies.  Data at \url{https://www.migrationpatterns.org}

Stelter, R. and Alburez-Gutierrez, D., 2022. Representativeness is crucial for inferring demographic processes from online genealogies: Evidence from lifespan dynamics. Proceedings of the National Academy of Sciences, 119(10), p.e2120455119.

\end{hangparas}

\pagebreak

\iffiguresInText
\fi

\ifforReview
    \setcounter{page}{1}   
    \singlespacing
\fi
\hspace{0pt}
\vspace{3cm}
\begin{center}
\fontsize{2cm}{2.4cm}\selectfont Supplementary Material
\break
\break
\normalsize Animated maps of measured and fitted parameters, as well as source code, may be found at \url{http://scaledinnovation.com/gg/migration/migration.html}
\end{center}
\pagebreak

\section*{SM1: Validation of genealogic data}

\noindent\textbf{\underline{Prior Validation}}  Our principal dataset has been extensively curated and validated by its authors (Kaplanis et al. 2018) and examined by others (Chong et al. 2022, Stelter and Alburez-Gutierrez 2022, Colasurdo \& Omenti 2024).  Principal criticisms involve small differences in mortality statistics in Scandinavia; our concerns are with location and distance, not age at death, and we principally examine North America with far higher sampling.\\

\noindent\textbf{\underline{Birth-Death Distance as Migration Distance}}
    At an individual level, migration distance is somewhat ambiguous: is it the single largest distance that someone may move in a lifetime?  Is it the summed magnitude of their moves, or the vector sum?  We have used genealogical birth and death locations to measure migration, which are appropriate as the literal beginning and end of a lifetime, as well as available in genealogies. Since the Kaplanis dataset includes (anonymized) parent and child identifiers, we can also examine waypoints, namely the births of any children.  The distributions below, for distances less than 4000 km (i.e. not trans-continental) are for parent birth to parent death distance (n = 828,000), parent birth to first-child birth (n = 854,000) and first-child birth to parent death (n = 893,000).  The only significant difference is that a smaller percent move from first-child birth to parent death ($f_{leave}$ = 0.55) than parent birth-death ($f_{leave}$ = 0.64) or parent birth to first-child birth ($f_{leave}$ = 0.62), as expected if most migration occurs before young adults have children.   In addition, the chart shows the typical quality of fit for the cumulative genealogic data (dots) to Equation 1 (solid lines).
    
\begin{figure}[H]
    \includegraphics*[width=8cm]{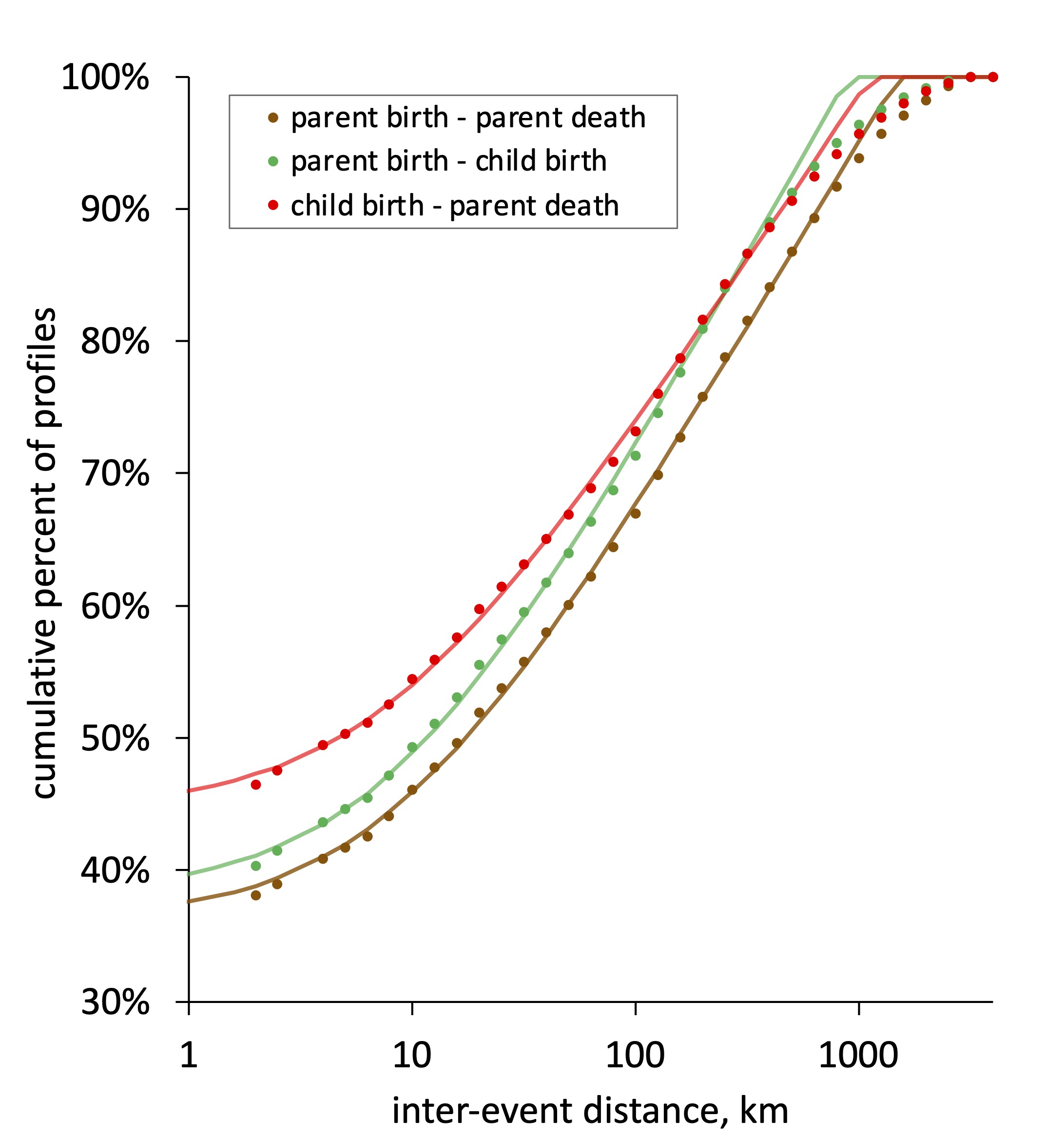}
    \centering
\end{figure}

The trends observed of subsets of birth-death distances, such as the $r_{fixed}$  spike in the US 1750-1890 or time-trends in $r_{max}$ (Figure 4) also appear in the parent-child distributions (not shown).\\

\noindent\textbf{\underline{Regional Bias}}
 Any regional bias — for example, if present-day Virginians disproportionately posted online family trees compared to Vermonters — is removed by averaging and fitting in smaller geographic blocks, then comparing and mapping across blocks.  Our conclusions are based on block-averages, normalized percents, and model fitting, not on absolute numbers.\\

\noindent\textbf{\underline{Time Bias}}
  The dataset contains more profiles of people born 1920-1950 than other generational slices, since online genealogy customers know their parents and grandparents better than earlier ancestors.  But as with geography, such bias is decreased by averaging and fitting to smaller blocks of time, and drawing conclusions from ratios and fits rather than absolute numbers.  All subsets used for mapping, averaging, or curve fitting contain at least 100 ancestor profiles.  We generally time-slice by thirty years to match generation times, which are also readily measured in the Kaplanis dataset: women = 29.7 ± 7.1 years, n = 762,000, and men = 34.4 ± 8.8 years, n = 916,000.\\

\noindent\textbf{\underline{Gender Bias}}
  Genealogic information will inevitably have a gender bias in a patrolynic culture: surnames do not provide continuity for female branches and ancestors, making them more difficult to trace.  Our gender-based analyses are therefore not across but within genders, e.g. SM2, SM3 below.\\

\noindent\textbf{\underline{Generic Location Error}}
  Kaplanis et al report birth and death locations as latitude and longitude which they compute by automated geocoding of user-entered place names.   In our context, a nation-scale descriptor, for example “Britain” or “Canada” is unacceptably vague, while a town or county level descriptor is useful.  We identify nation-scale profiles as very high-density blocks in central unpopulated areas, and ignore them in our tallies.\\

\noindent\textbf{\underline{Specific Anomalies}}
  Any crowdsourced genealogy will contain anomalies; algorithms can identify only some types of error, i.e. children older than parents.  In Kaplanis we find, for example, reported ancestry in Oregon in the early American colonial era, which is either simply incorrect or may indicate an estimate of Native American ancestry.  Rather than impose judgment, we generally require all sampled time-and-space data subsets to contain at least 100 profiles and use robust measures, median or geometric mean, that are less sensitive to outliers especially with a power-law distributed measure like migration distance.\\

\noindent\textbf{\underline{Comparison to Census Results}}
  The center of population, computed from census data, is an iconic measurement in demographics (\url{https://www2.census.gov/geo/pdfs/reference/cenpop2020/COP2020_documentation.pdf} ).  It may also be estimated from genealogies: we count everyone alive in each tally year (each decade, to match US census data), and note their location as their birth location if under age 20 and death location if older.  The result, shown below, is a time-track for the US center of population that extends back to the early colonial era, well before the first census in 1790.  The orange ellipses extend one standard deviation in latitude and longitude from each center point and show that the genealogic and census population centers coincide, well within one sigma, from 1800 to 1950, after which, as noted elsewhere, genealogic data become very sparse. 
   
\begin{figure}[H]
    \includegraphics*[width=12cm]{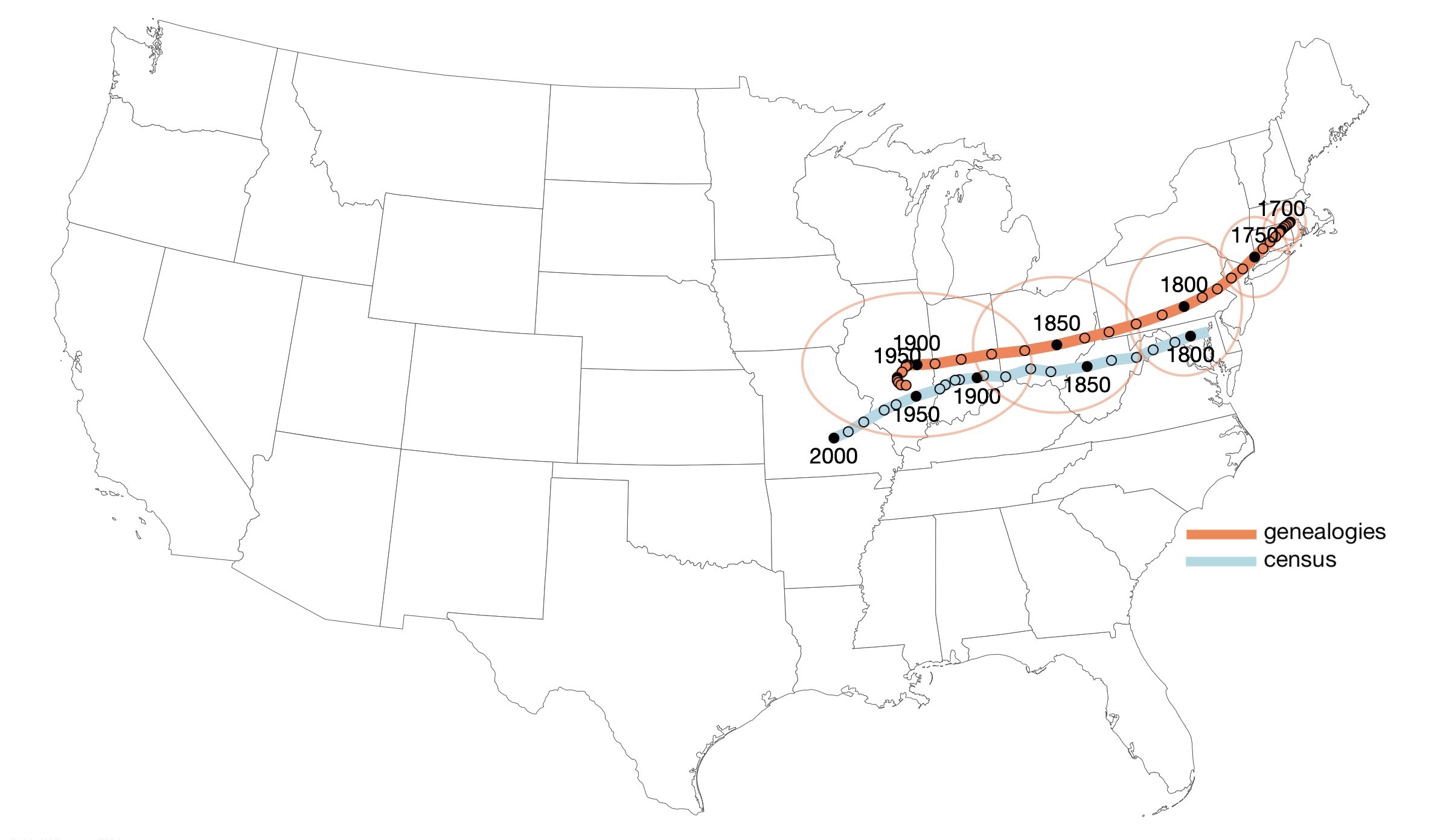}
    \centering
\end{figure}
   
This is a stringent comparison since it requires an estimate of total population (rather than percent or model fit) and is therefore subject to any regional bias.  The systematic 0.7° latitude northerly difference between genealogic and census centers (1800-1900) probably reflects such a bias, which is consistent with the much greater density of historic societies north of this line (Rosenstein \& Vakharia 2022, Figure 4) which support a deeper record-keeping of genealogic information.

Many other maps are available; in particular the movement of the frontier, as measured by population density (e.g. \url{https://www.census.gov/dataviz/visualizations/001/}), is also illustrated by the transition from migration to settlement \url{http://scaledinnovation.com/gg/migration/migration.html?type=origin&loop} .

\pagebreak
\section*{SM2: Dependence on Gender}

The chart below shows distributions by gender for 1.5 million birth-death distances  in three subsets (Britain and Ireland (all dates), female n = 174,143, male 239,192; American 1620-1770 female 216,023, male 269,400; American 1770-1860 female 255,475, male 354,906). Least-squares fits to Equation 1 are shown as solid (American) or dashed (British-Irish) lines.  Any gender differences are small compared to the differences that interest us, namely those across time or geography.

\begin{figure}[H]
    \includegraphics*[width=10cm]{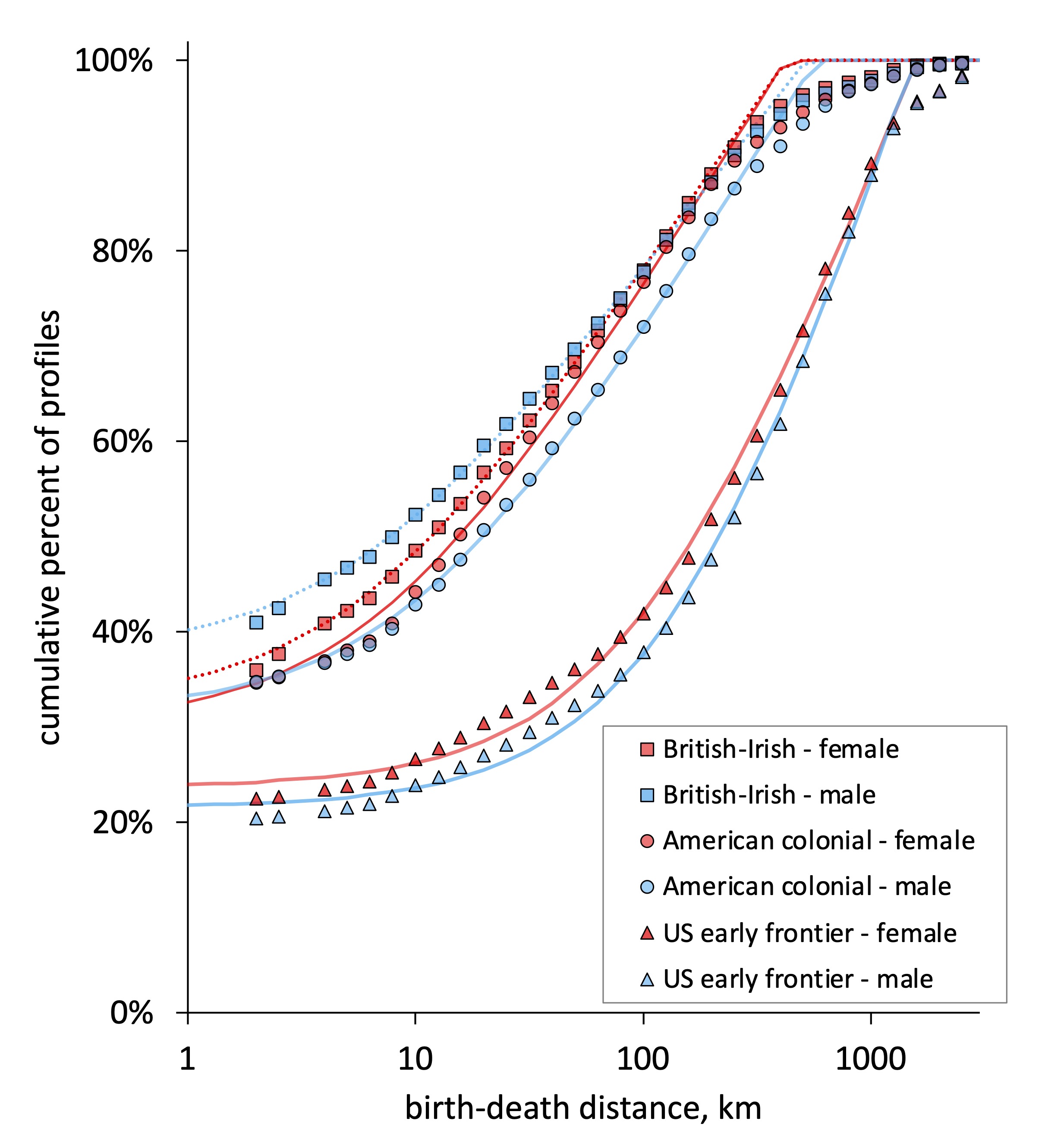}
    \centering
\end{figure}

\pagebreak
\section*{SM3: Age at migration}

The age at which people move is relevant to a discussion of their migration, but is not directly available from genealogic data that record only birth and death.  But since the Kaplanis data include parent-child links (via anonymous id numbers), we have dates and locations for intermediate life-events, namely the births of children, which we analyze in three ways.
A child born before their parent’s migration will be closer to that parent’s birthplace than death-place, and vice-versa.  The chart below shows geometric means of the distance between genealogical events for 1.66 million profiles, all regions and dates.  The crossover point at which the (parent birth)-(child birth) distance equals the (child birth)-(parent death) distance should occur at the parent’s age of migration; this occurs at about age 18 for women and age 23 for men, consistent with migration in early adulthood.

\begin{figure}[H]
    \includegraphics*[width=10cm]{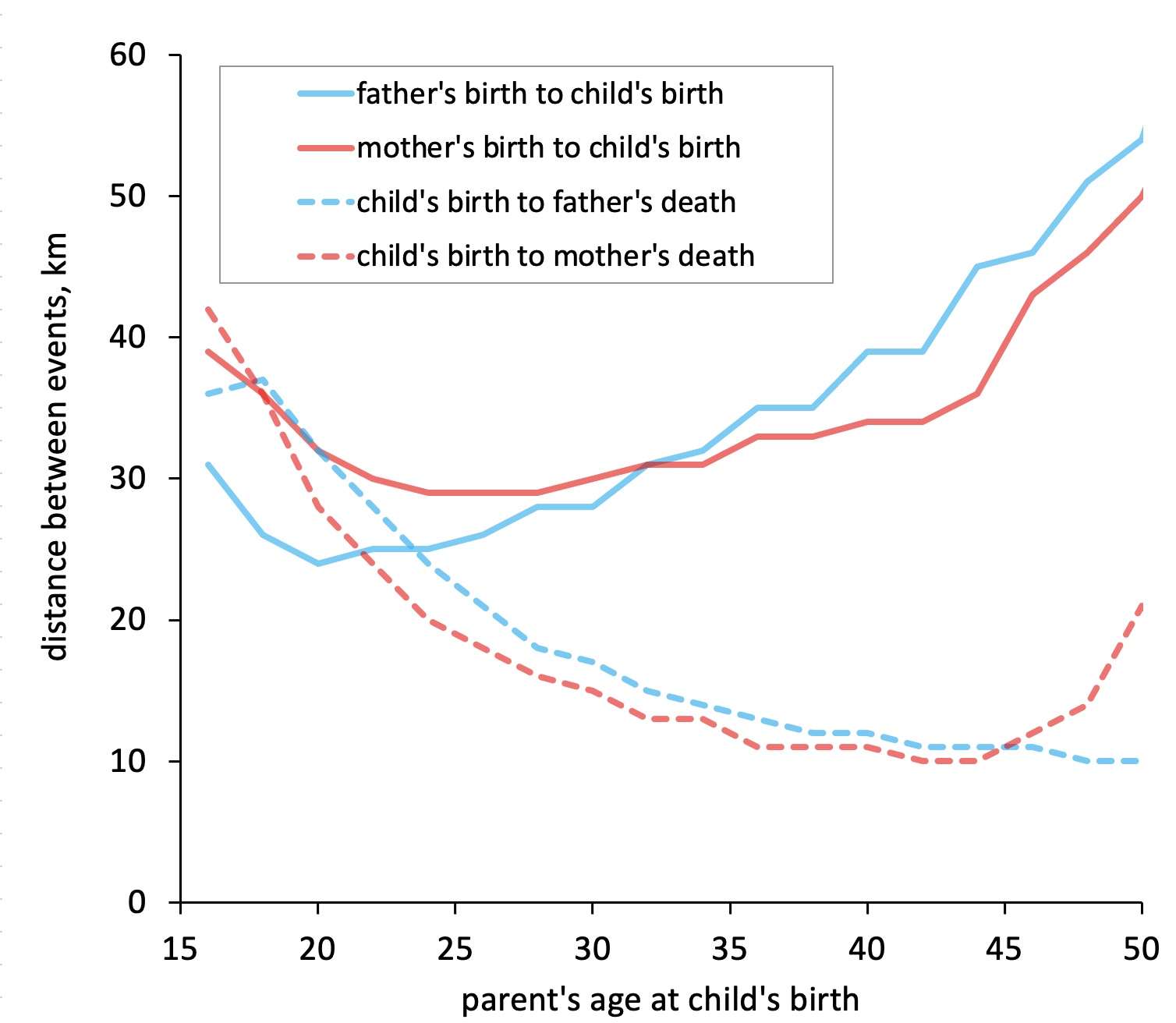}
    \centering
\end{figure}

Births of multiple children are milestones along a parent’s life-path. The results below  are from 27,800 profiles of American internal migrants, all with four or more children constrained to a near-linear path from parent birth, to all child-births, to parent death. These data are generally consistent with Rogers’ consensus age-of-migration model (Rogers \& Castro 1981, table 17, shown below as its cumulative distribution) which peaks at age 23, though the peak age in early frontier era (1770-1860) is about 10 years later than in the colonial era (1620-1770) or later period (1860-1950).

\begin{figure}[H]
    \includegraphics*[width=8cm]{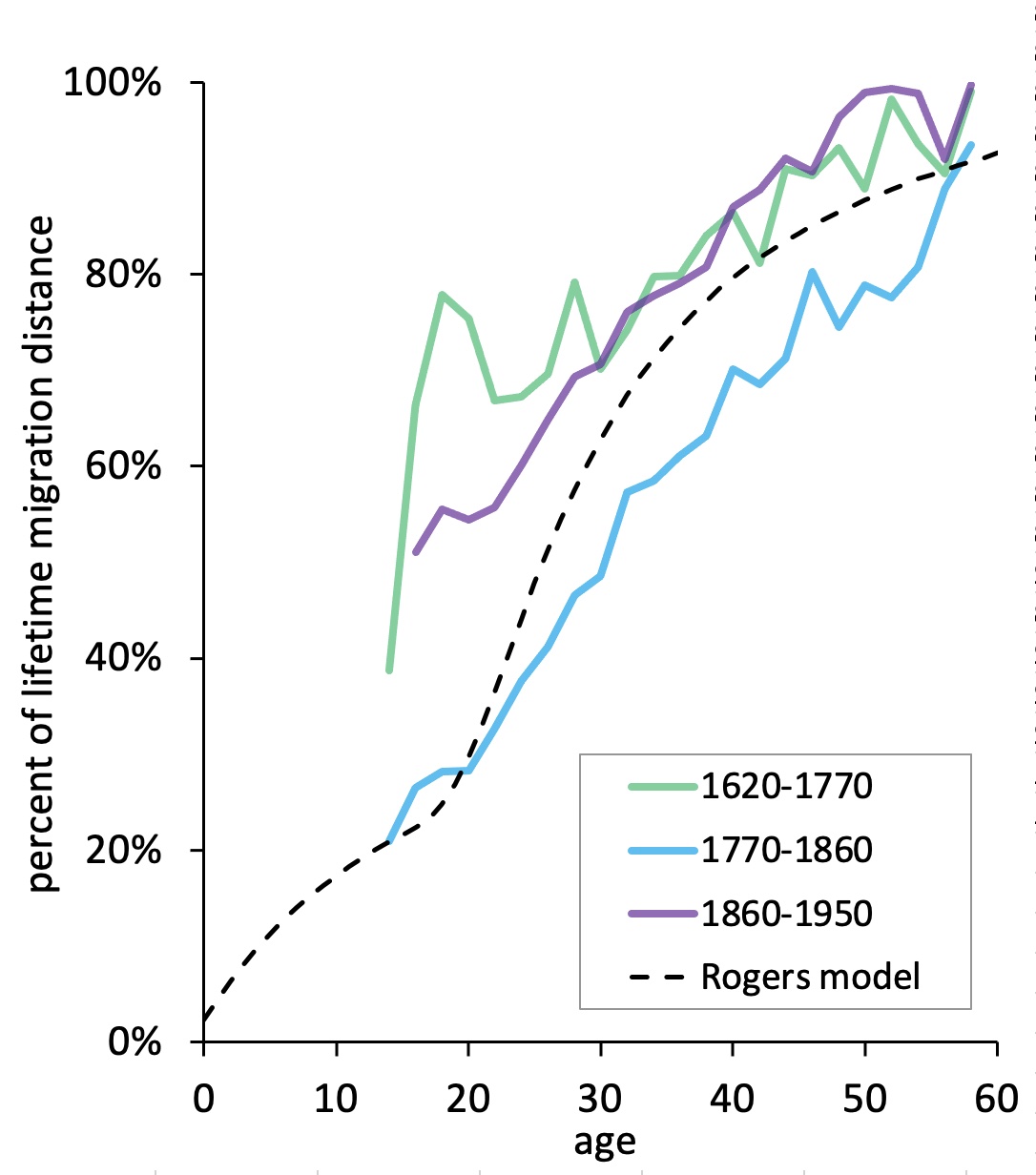}
    \centering
\end{figure}

Thirdly, as noted above (SM1, Birth-Death Distance as Migration Distance), the distance distributions between a parent’s birth, death, and their first child’s birth, are also consistent with migration in young adulthood.

\pagebreak
\section*{SM4: Three-parameter model}

Based on a Zipf-like view that migration cost is proportional to distance and the simplicity of the distributions of Figure 1, the basis for any mathematical description is \\~\\
\indent$p_{migrate} \approx 1/r$\\~\\
“Cost” could also include a linearly adjustable distance-independent component\\~\\
\indent$cost \approx (1 - f) + f\cdot r ; f = 0…1$  \\~\\
When f → 0, cost → 1, and when f → 1, cost → r.  Adding an independent stay-leave parameter (because only the leavers pay a moving cost) and replacing $r$ with this cost function gives \\~\\
\indent$p_{migrate} \approx (1 - f_{leave}) + f_{leave}/(1 - f + f\cdot r)$ \\~\\
Integrating and substituting $r_{fixed}$ = 1/f to make the logarithm scale-free and cast the fixed-cost parameter into distance units, gives \\~\\
\indent$CDF \approx f_{leave} ln( 1 - 1/r_{fixed} + r/r_{fixed} )$ \\~\\
Removing the scale dependence inside the logarithm and making the assumption that interesting distances are $>>$ 1 km gives  \\~\\
\indent$CDF \approx f_{leave} ln( 1 + r/r_{fixed} )$ \\~\\
Normalizing so that CDF → $f_{stay} = 1 - f_{leave}$ as r → 0 and CDF → 1 as r → $r_{max}$  gives \\~\\
\indent$CDF \approx 1- f_{leave}(1 - ln(1 + r/r_{fixed})/ln(1 + r_{max}/r_{fixed} )) $ \hfill (1)

\begin{figure}[H]
    \includegraphics*[width=7cm]{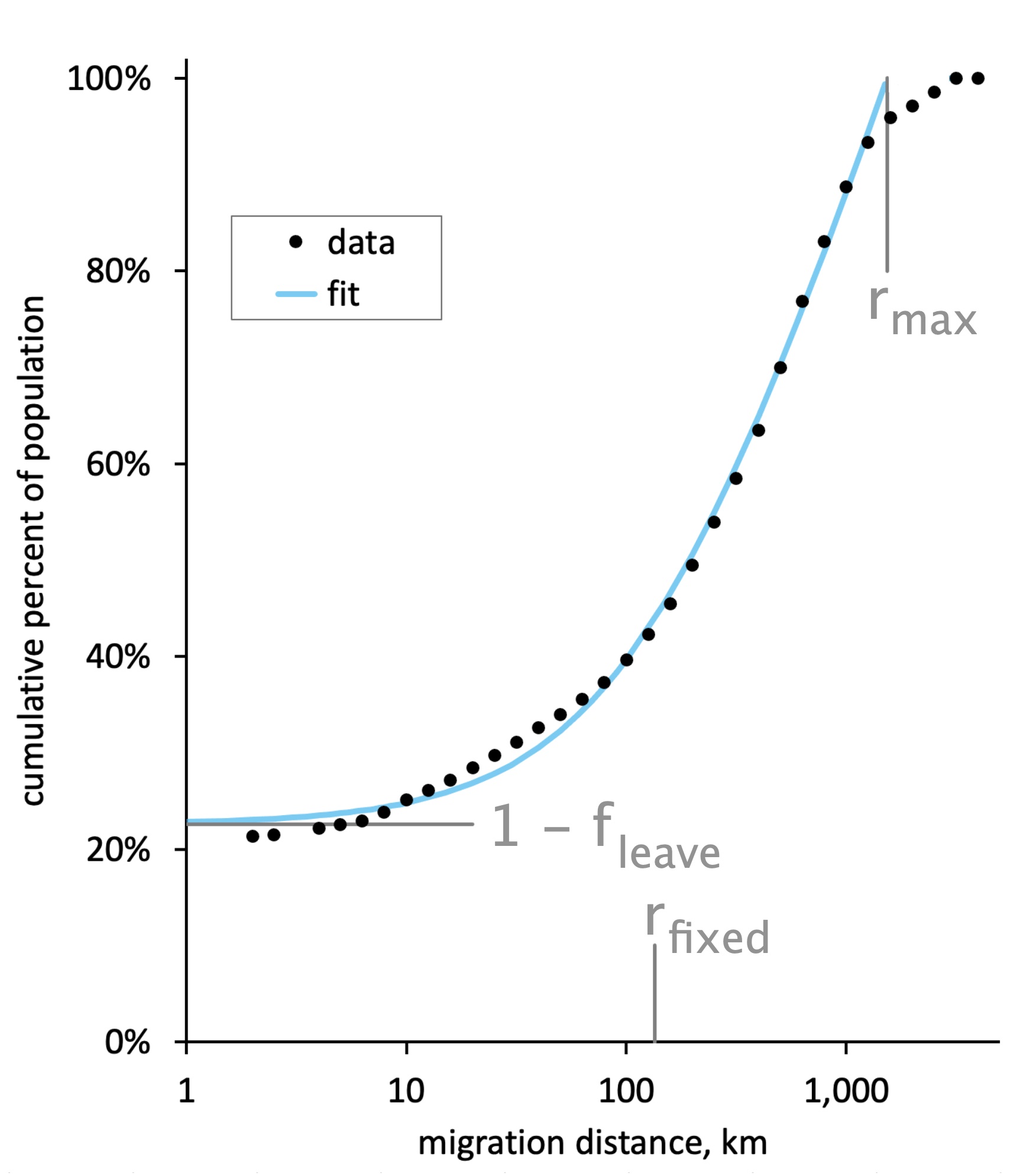}
    \centering
\end{figure}

The interpretation of these parameters is shown above with the least-squares fit to a typical dataset (US 1770-1860; 608,000 profiles), in this case with fitted parameters $f_{leave}$ = 0.774 ± 0.002, $r_{max}$ = 1532 ± 27 km, $r_{fixed}$ = 135 ± 5 km, rms error 1.3\%. Other algebraic forms (additive, multiplicative, reciprocal, etc.) were tried for a cost parameter with poorer fits and greater systematic differences from the observed data. Equation 1 is justified by its simplicity, facile interpretations for $f_{leave}$ and $r_{max}$, and fits to observed data. It is not derived by analogy to inanimate physics.

\pagebreak
\section*{SM5: Model Parameter Independence}

Even though our model has only three fitted parameters, it is important to see if they are correlated.  We created a  test dataset by partitioning North America into one degree latitude by longitude blocks, then tallying all profiles for each block for those age 20 within each 30-year time slice from 1620 to 1950.  Only subsets with at least 100 profiles were kept (n = 2577) and their birth-death distances fitted to Equation 1.

The parameters are uncorrelated (graphs below: top row).  Median birth-death distance is a robust observable (not a fitted result) and is correlated with all three fitted parameters : $r_{fixed}$ (strongly), $r_{max}$ and $f_{leave}$ (less so) (bottom row). 

\begin{figure}[H]
    \includegraphics*[width=12cm]{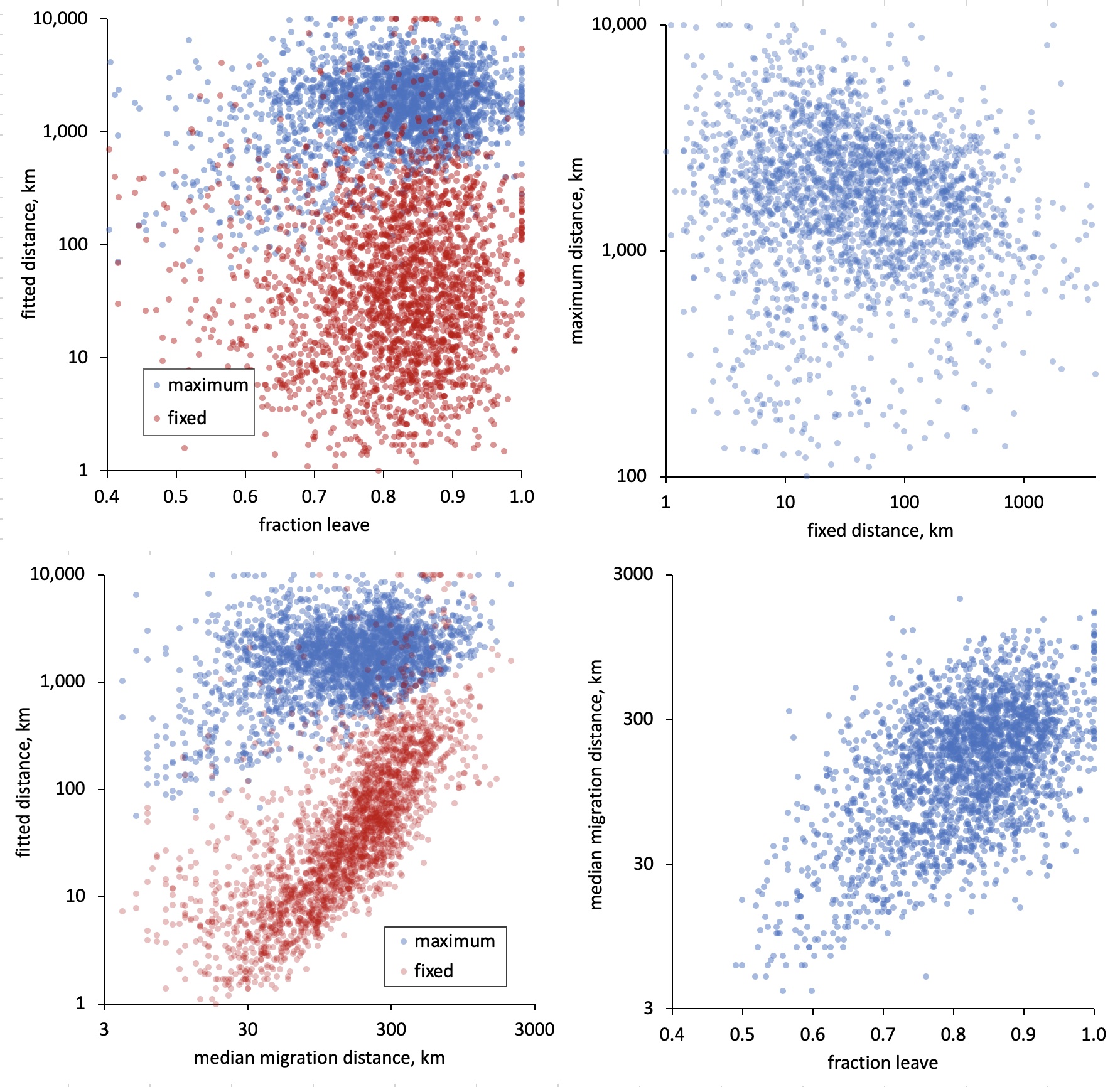}
    \centering
\end{figure}

\pagebreak
\section*{SM6: Standardizing Migration Data}

The diversity of sizes of administrative areas has long frustrated attempts to compare percents migration across countries and epochs (Long \& Boertlein 1976).  When the distance between administrative areas is large, the underlying 1/r distribution of migration distances results in an underestimate of those moving.  Viewing migration distances as a distribution and fitting that distribution to a model, e.g. Equation 1, can reduce the bias in the simplest migration question: How many people moved?

Two datasets can be compared if they have the same resolution limit, which is the smallest accurately measurable distance.  For the Kaplanis genealogy data, that limit is, empirically, 10-20 km = $r_{fixed}$ for places and time periods except the clear American anomaly 1760-1890 (Figures 4, 5).  This is consistent with a simple estimate of the average distance between towns:  contiguous US area $\approx$ 8,000,000 km$^2$,  20,000 incorporated villages, towns, and cities gives 40 km$^2$ per town, thus $\sqrt{40} \approx 6$ km between neighboring town centers.  In other words, the observed resolution coincides with the expectation that genealogic event recording is at the scale of towns and villages.  Most migration studies depend on data collected by administrative area — county, state, or country — in which case the resolution limit is set by the distance between those areas.  Given the 1/r distribution, in which there are many more short jumps than long jumps, an average distance between the centers of nearest-neighbor administrative areas is a a good measure of the resolution limit.

Datasets with different resolution limits can be compared either by data truncation or by distribution extrapolation. In truncation, migration distances for the finer-grained dataset below the longer limit are set to zero; i.e. if the coarser limit is 100 km, a person who moves 75 km is judged not to have moved at all.  Extrapolation involves extending the 1/r trend (= log trend in the CDF) of the coarse dataset down to the limit of the fine-grained dataset.  Both methods agree for a comparison of the US 2010 census data (tallied by commuting zones with a resolution limit $\sim$80 km) to the Kaplanis data, and show that young-adult tendency to leave home in 2010 was comparable to what it was in 1710, which is to say 50-60\% instead of the uncorrected 30\% (SM7).

\pagebreak
\section*{SM7: 2010 US Census Migration Data}

Since the genealogy-derived Kaplanis data do not extend much past 1950, any data that illuminate recent behavior is of interest. Recent US data are available for 31 million young adults in 2010 (Sprung-Keyser et al. 2022), and as shown in Figure 2a, also display the 1/r distribution of migration distance.  The CDF is shown below-left, overlayed for comparison with the Kaplanis data from 1710, 1800, and 1950 which are also shown in Figure 2.

The 2010 data measure the locations of people born between 1984 and 1992 as tallied in 2000 and 2010 census and taxation data, essentially measuring where they were at age 16 and then at age 26.  This is not the birth-to-death distance of the Kaplanis data but should capture the peak years for migration.

The 2010 data have a fitted value of $r_{max}$ of 2380 km, at the plateau since 1830 (Fig 4b) and which simply reflects the maximum possible migration distance (the numbers moving between the contiguous states and Alaska or Hawaii are small).  The vertical intercept, $f_{stay}$ = 1 - $f_{leave}$ , is 70\%, well above the 20-35\% range of the Kaplanis datasets, but we do not think that this implies that people stay close to home much more in recent years than in earlier decades.  Instead, this reflects the distance resolution problem when comparing different data sources.  Kaplanis locations are latitude and longitude from geolocation, and therefore largely at town-level resolution. Sprung-Keyser is coarse, placing a person in one of 741 commuting zones (CZ); as shown below-right, the mean distance from the geographic center of a CZ to its nearest neighbor is 83 ± 44 km, exactly at the sharp break in the CDF (gray dots, below left).  Simulated truncation of continuous data, for example setting all distances below 80 km to zero, gives the same sharp transition from a 1/r distribution to a flat plateau.

Extrapolating the 1/r region from ca. 80 km to 1 km resolution (gray dashed line, below left) suggests that a value of $f_{stay}$ of about 40\% is more comparable to the Kaplanis datasets; in other words modern stay-vs-leave behavior has reverted a range typical of pre-industrial Europe or pre-frontier America.  Truncation gives the same result, since the census data (gray dots) overlaps the colonial distribution (light blue) at the former’s resolution limit of 80 km.

\begin{figure}[H]
    \includegraphics*[width=12cm]{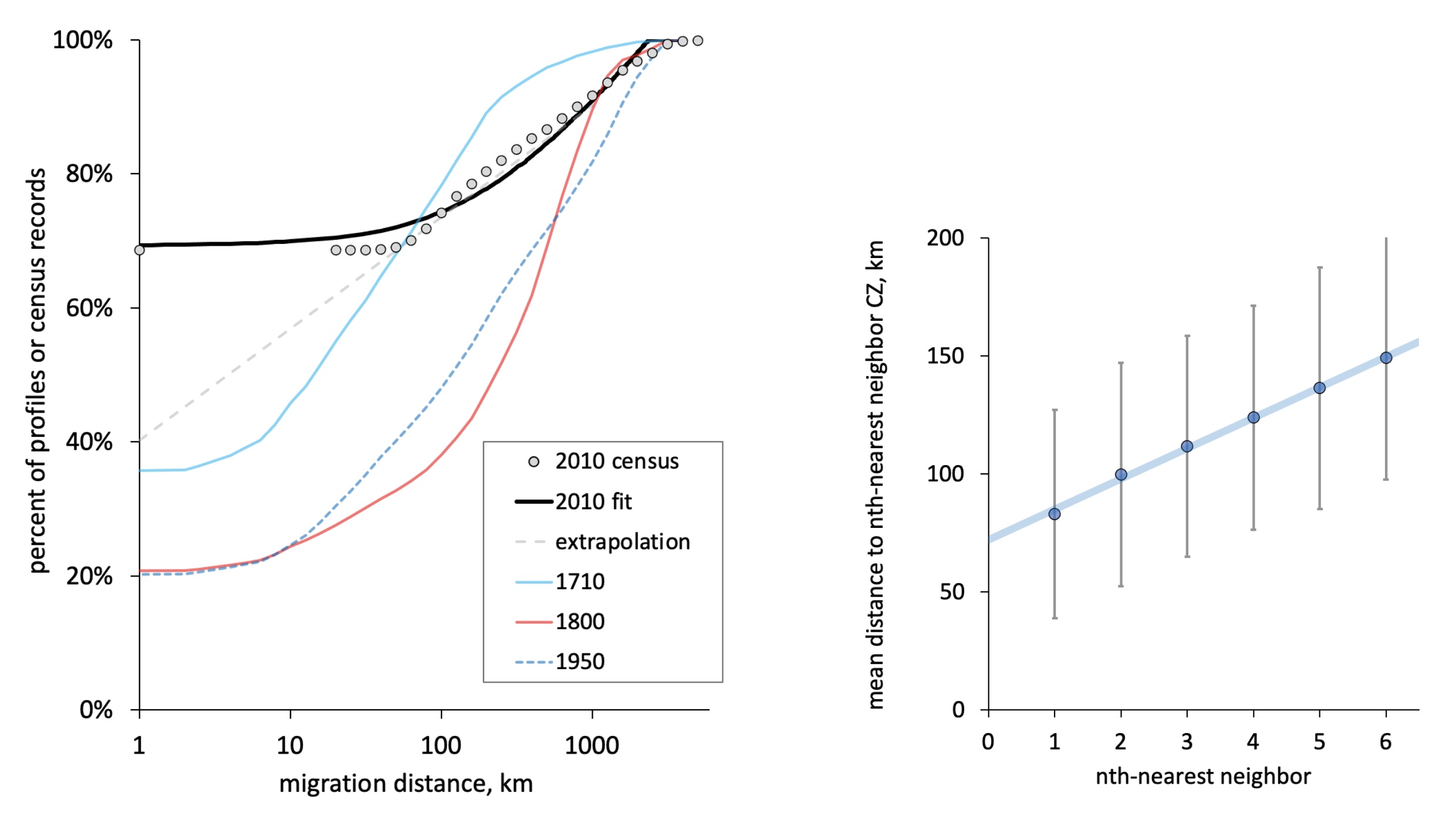}
    \centering
\end{figure}

\pagebreak
\section*{SM8: Geographic Dependence of the Fixed Distance Parameter}

The $r_{fixed}$ term in Equation 1 displays a unique spike in the American data from 1770-1860 (Figure 5), with values of 50-250 km that cannot be explained as a resolution limit in the dataset.  The map below-left shows the median values of fitted $r_{fixed}$ in this era for every 0.5°x0.5° block with at least 100 profiles. A geographically distinct pattern is apparent with low values on the New England coast and very high values in western New York and inland Virginia to South Carolina.  The lower chart shows the data (dots) and fits (lines) for the lower and upper third of these geographic blocks; the distribution for the low $r_{fixed}$ blocks (blue) is very similar to that for all colonies 1620-1710 (Figure 3) with a higher $r_{max}$.

\begin{figure}[H]
    \includegraphics*[width=12cm]{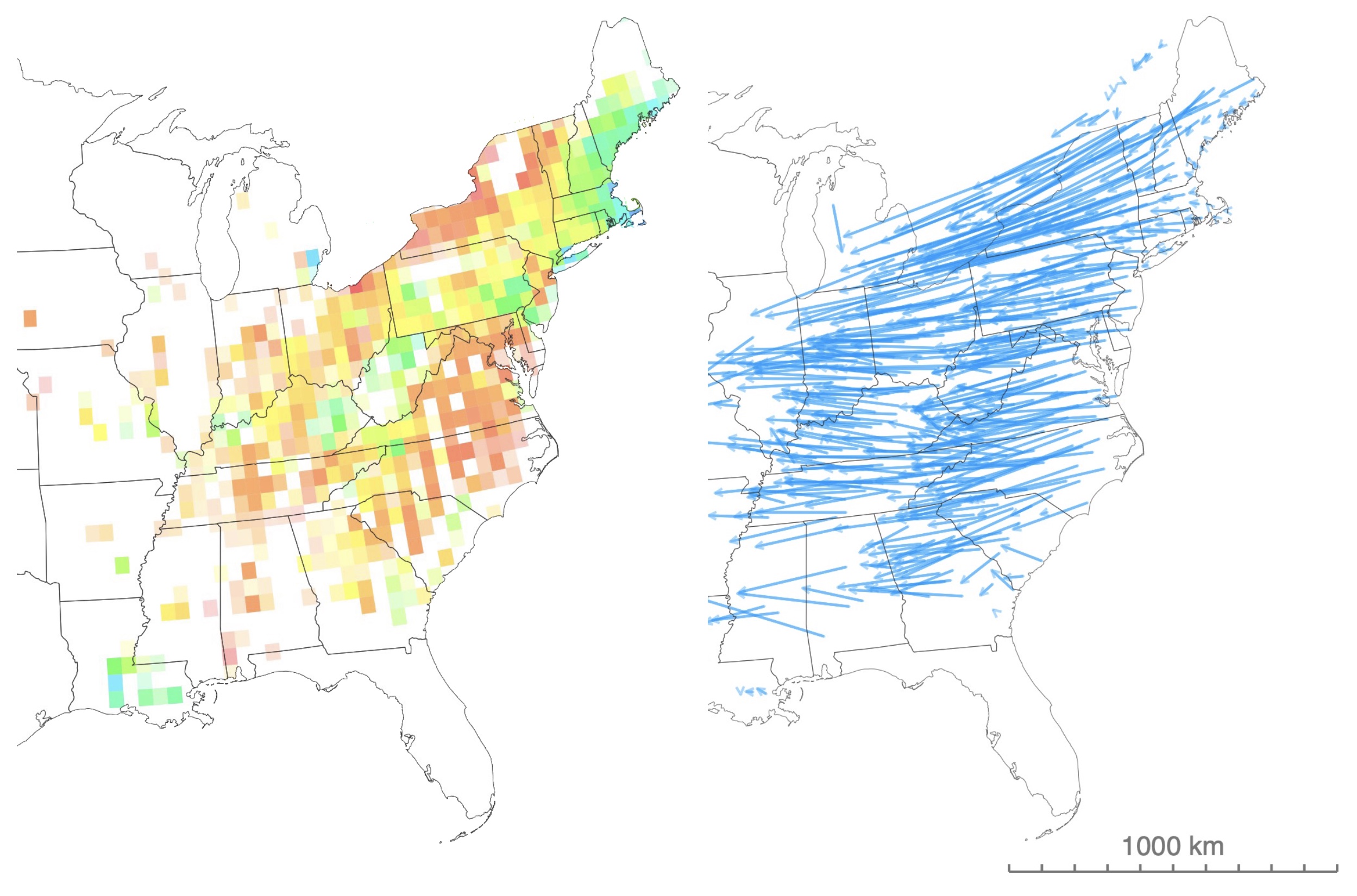}
    \centering
\end{figure}

A physical explanation is suggested by the vector sums of all migrations for the blocks (above-right): most migrations from the high-$r_{fixed}$  blocks have infeasible or undesirable barriers to the west, either Ontario or a Great Lake, or the Appalachian Mountains.  With reports that better land was available across those barriers, the decision to migrate may have had an implicit minimum fixed distance.

\begin{figure}[H]
    \includegraphics*[width=8cm]{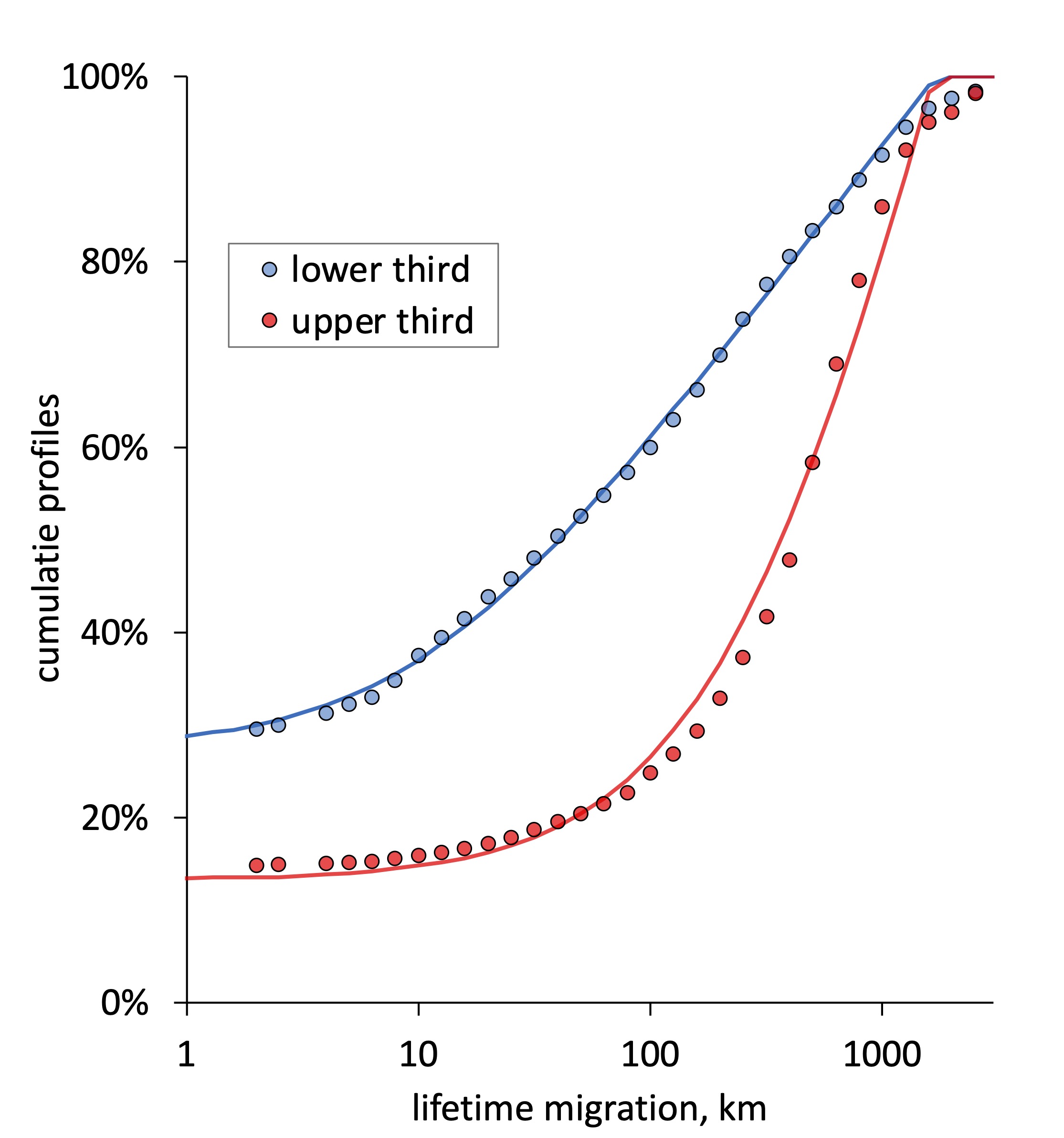}
    \centering
\end{figure}

We note that in a closed system, the combination of high $r_{fixed}$  and $f_{leave}$  would be self-extinguishing: most children would move a considerable distance away from home.  American internal migration away from the east was, however, back-fueled by a very high net survival rate as well as high transatlantic immigration.

\pagebreak
\section*{SM9: Directionality of American Migration}

The directionality of American internal migration 1620-1950 is shown below.  Each blue dot is the bearing (0 = North, 90 = East, etc) of the vector sum of all birth-death vectors for all 0.5° square geographic blocks in North America in the given 30-year slice, with a minimum of 100 profiles.  Dots are horizontally dithered to show their distribution; orange lines are the mean ± standard deviation of the bearing for each time period.
\begin{figure}[H]
    \includegraphics*[width=12cm]{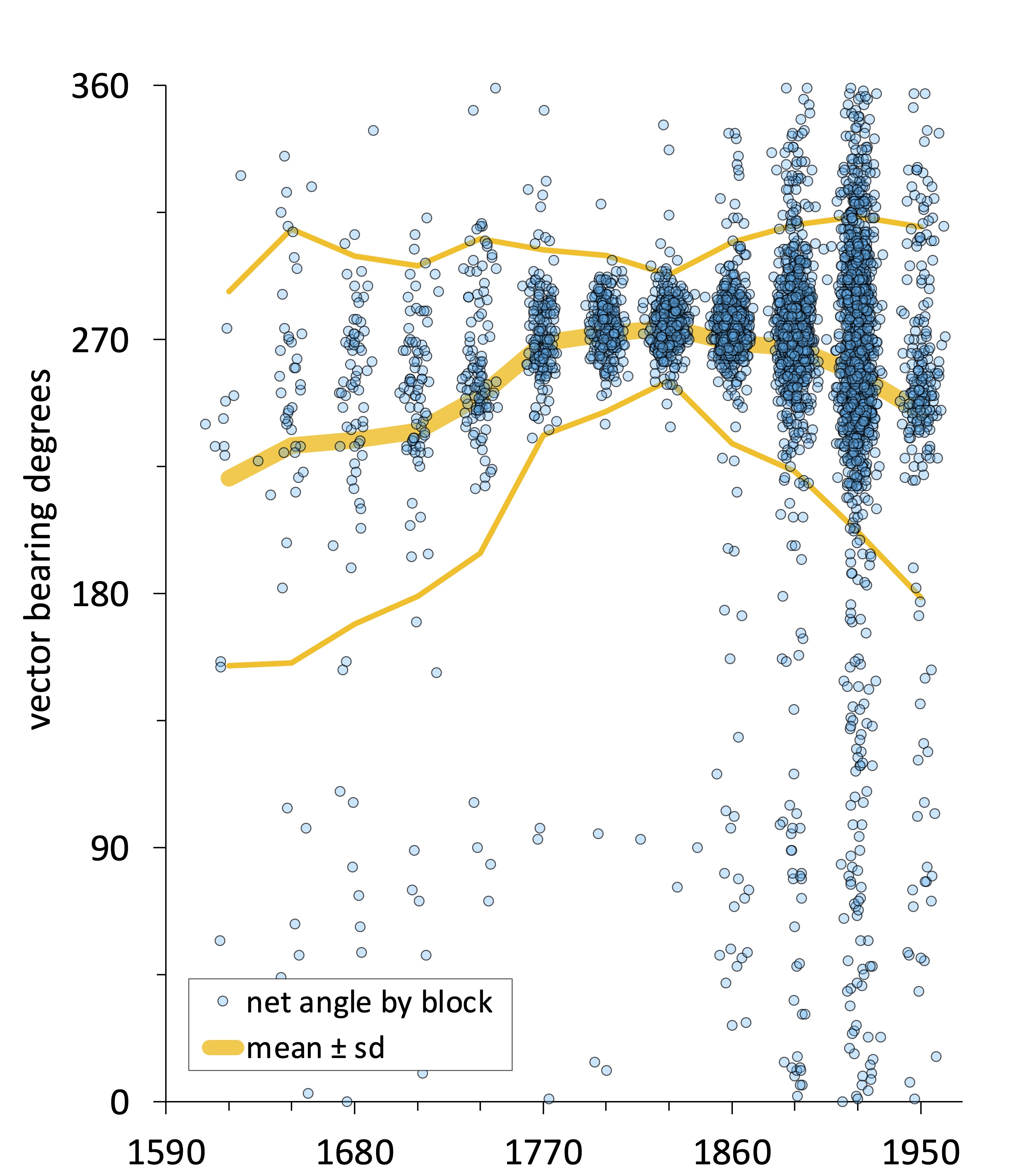}
    \centering
\end{figure}

\pagebreak
\section*{SM10: Migration Symmetry}

Migration symmetry $S$ is computed for a geographic block (typically a county or latitude-longitude block) as \\~\\
\indent $S = (births - deaths)/(births + deaths)$  \\

\begin{wrapfigure}{r}{0.25\textwidth}
    \includegraphics[width=0.24\textwidth,trim={1cm 2cm 0 5cm}]{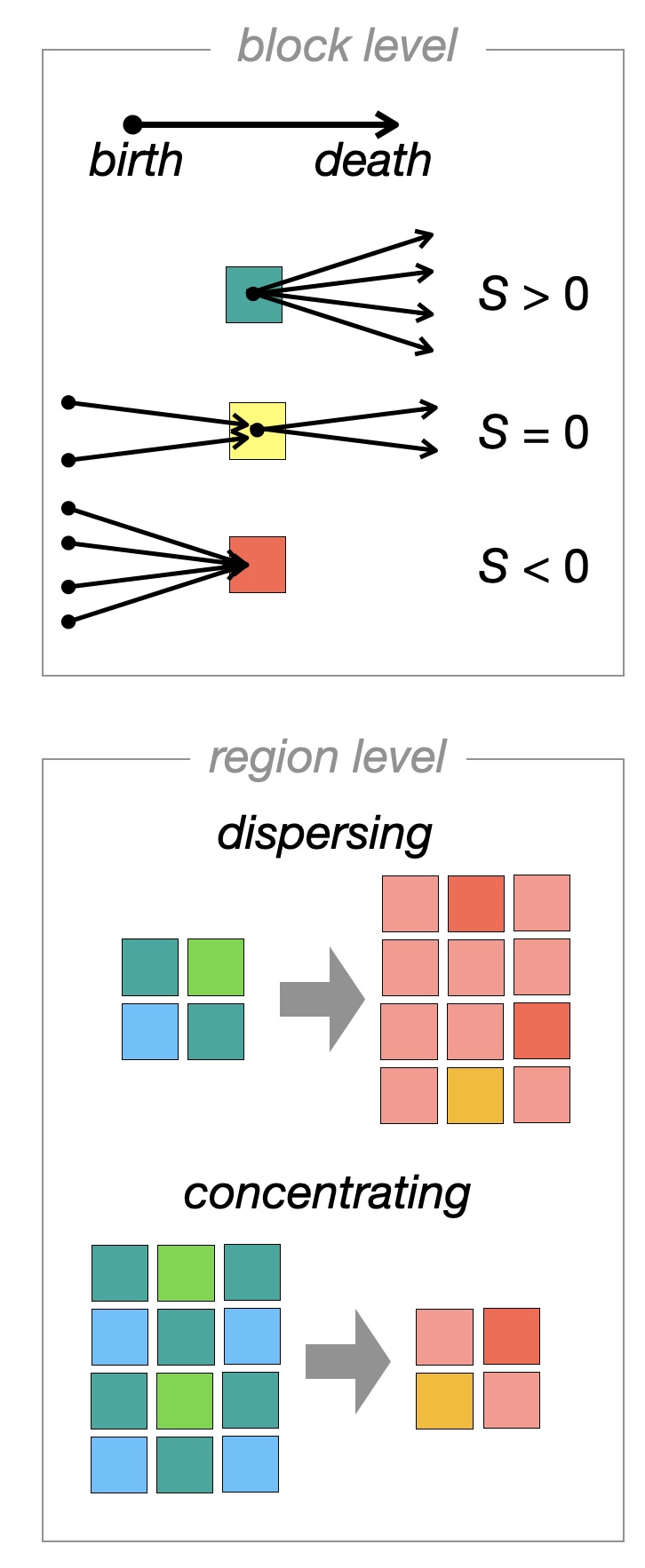}
\end{wrapfigure}

\noindent where births = number of genealogic profiles with birth location in the given block but death outside, and deaths = number born outside the given block but died within.  Symmetry is a migration metric, thus within-block births and deaths are ignored.  If S equals zero, people may be moving but influx equals efflux and the block is at a migration steady-state. The red-to-yellow-to-blue scale at the right is the same used in the maps below.

At larger scale, if a few blocks lose people to a larger number of blocks ($\Sigma S < 0$), the population is dispersed and diluted.  This is typical of an early frontier in which the destination area is larger than the source area.  Urbanization is the opposite, in which a larger number of usually geographically disperse blocks feed a smaller number and the population becomes concentrated ($\Sigma S > 0$).

Just as for the distance metrics, American migration is well described by both distributions and maps of symmetry over time. 

The distributions (below; left:data, right:illustrative sketch) are of the means of S for all 0.5°x0.5° blocks with $births + deaths > 100$ (number of blocks n=25 for 1620, n=1717 for 1920).  Until $\sim$1680, the colonies were close to a steady state with a centrosymmetric distribution of $S$.  The 1710 map (see SM11) shows the first development of a frontier, apparent in the distributions as a sharp rise in blocks with $S \approx -1$, i.e. unidirectional migration into new territory, a trend which continues for 5-6 generations.  
\begin{figure}[H]
    \includegraphics*[width=10cm]{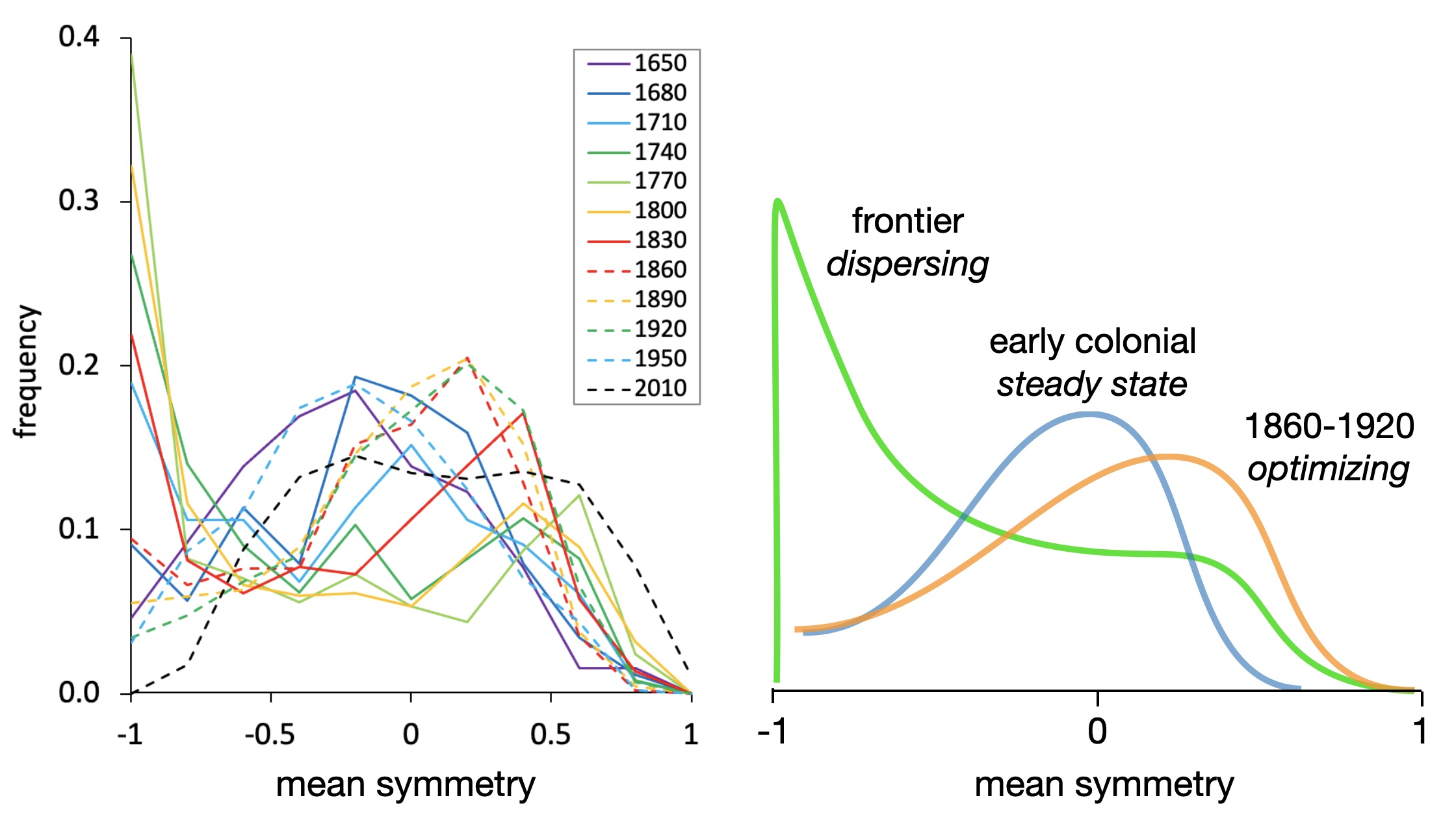}
    \centering
\end{figure}
By 1830 the distribution becomes bimodal with a preview of western urbanization (Chicago, Salt Lake City, San Francisco, and Los Angeles). For 1860-1950 a steady-state dominates east of the Mississippi (yellow and light green) and the pronounced asymmetries are urbanization and moves to California and Florida.  Census data for 2010 commuting zones (Sprung-Keyser et al 2022) show a return to a centered distribution but with a mean absolute symmetry of 0.38, still very dynamic.

While non-zero symmetry $S$ highlights different situations, frontier migration is distinguished by being highly directional, as shown by the strong east-west color gradient 1710-1830.  Urbanization pulls from diverse locations and is not uniquely directional.

\begin{figure}[H]
    \includegraphics*[width=12cm]{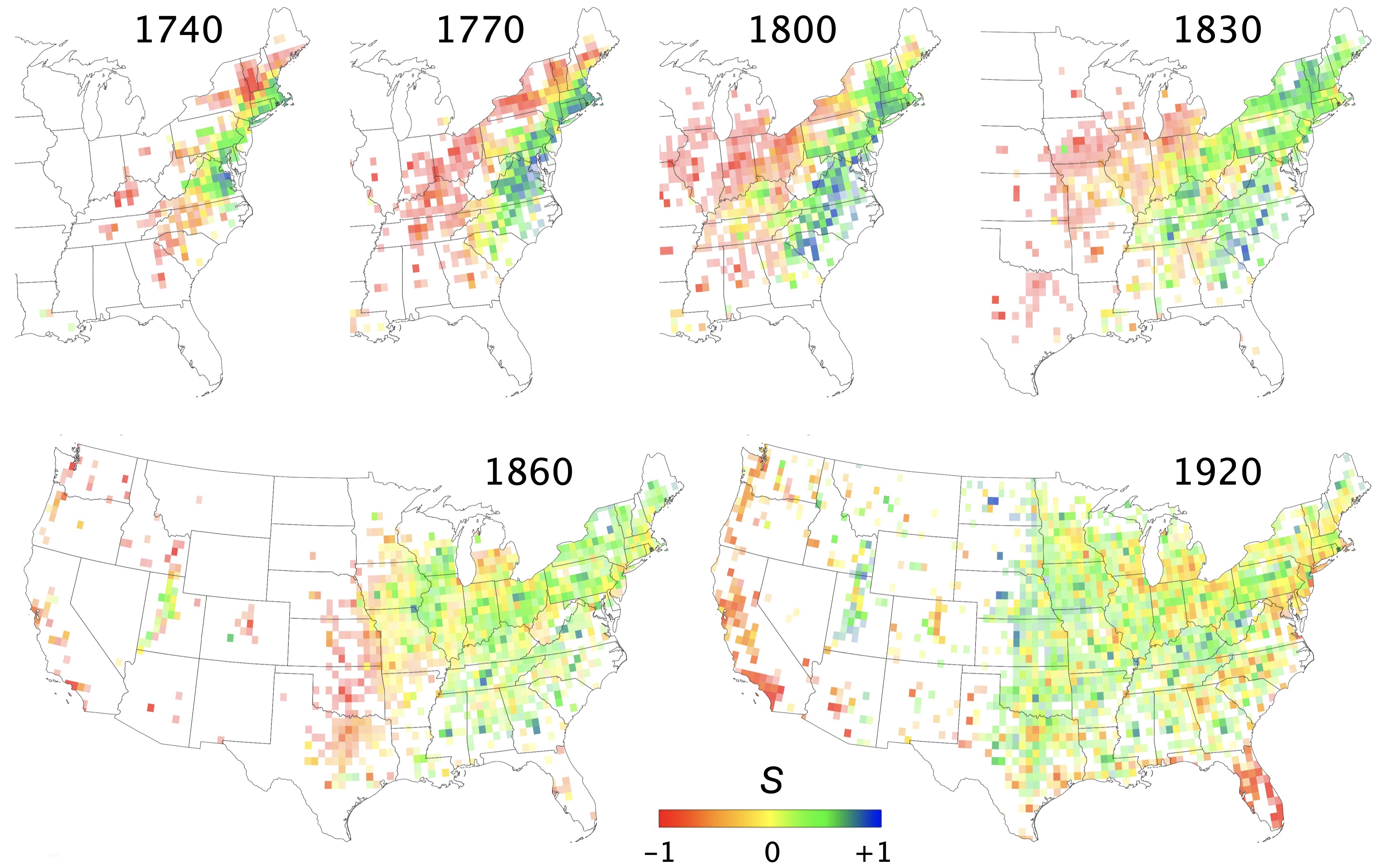}
    \centering
\end{figure}

\pagebreak
\section*{SM11: Alternative Frontier Mappings}

The American frontier is classically depicted as contours of population density based on decennial censuses since 1790, such as (a) Gannett’s map for 1830 (Gannett 1903) and its online equivalents at \url{https://www.census.gov/dataviz/visualizations/001/}.  We offer two alternatives from genealogic data.

\begin{figure}[H]
    \includegraphics*[width=12cm]{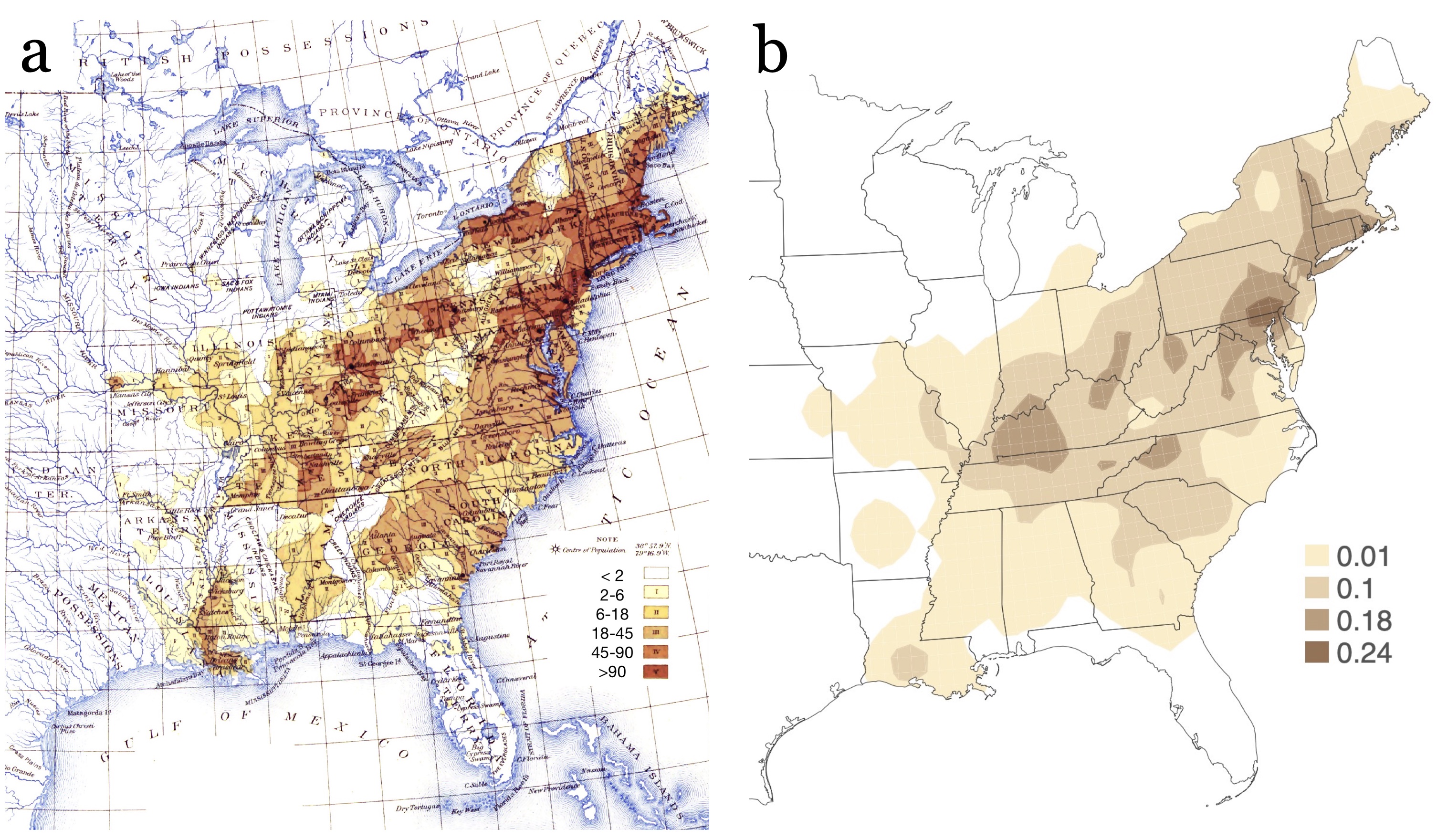}
    \centering
\end{figure}

\begin{wrapfigure}{r}{0.4\textwidth}
    \includegraphics[width=0.38\textwidth,trim={0 0 2cm 2cm}]{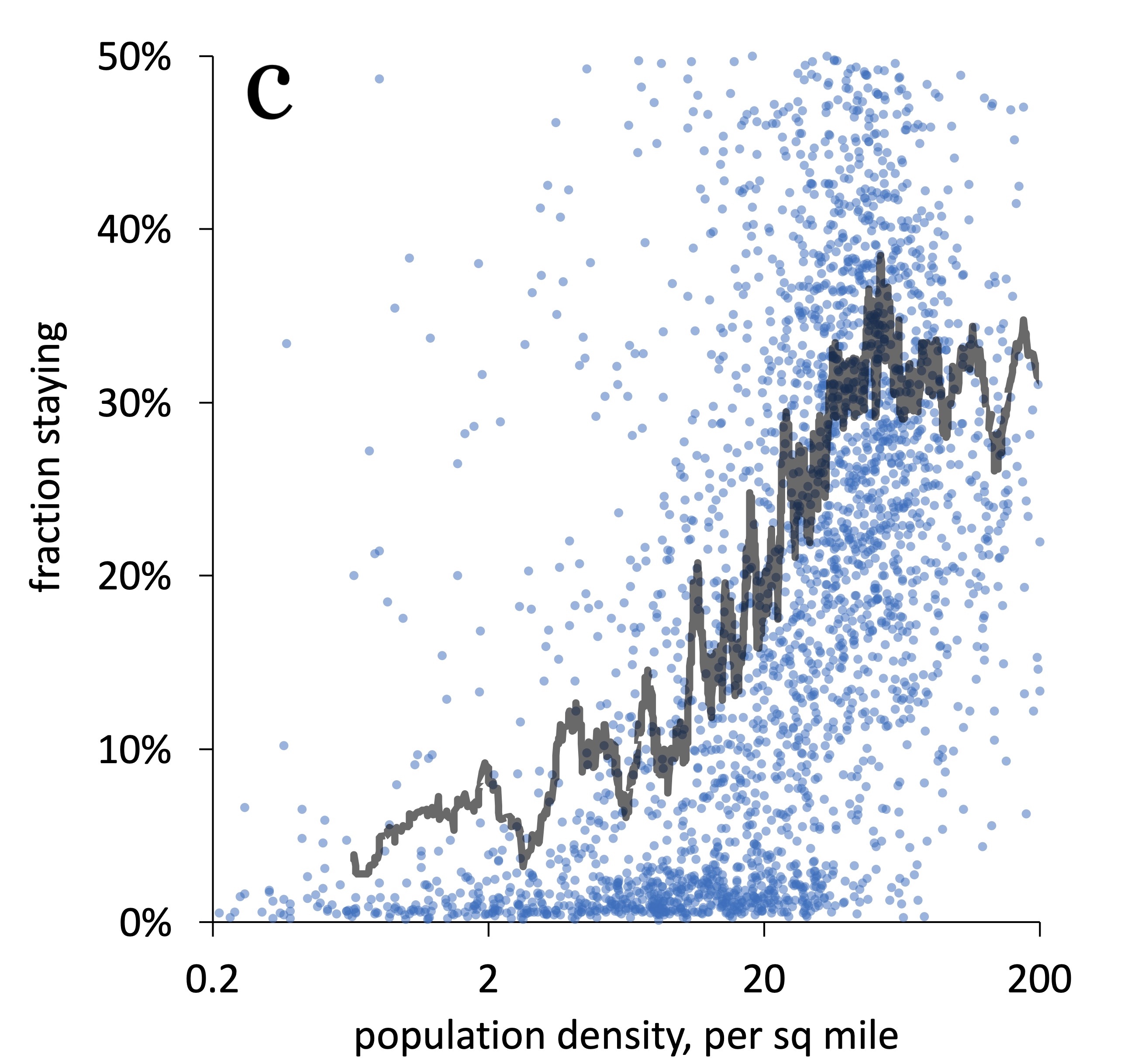}
\end{wrapfigure} 

A reasonable match (b) is seen with contours of $f_{stay} = 1 - f_{leave}$, obtained by fitting Equation 1 to all profiles in each 0.5x0.5° block. 

Chart (c) shows the relation between population density and $f_{stay}$ for 3200 counties’ census data 1790-1890 and 100 or more genealogy profiles per county in each decade.

The relation is monotonic and shows a rapid rise in “staying” behavior at 6-20 people per square mile, albeit with considerable scatter (the dark line is a smoothed running average).  The traditional Gannett frontier line corresponds to this rise, i.e. the first few percent of those who don’t move.  Like population density, $f_{stay}$ is a threshold metric: it does not revert but generally keeps rising as the frontier passes by.

\begin{wrapfigure}{r}{0.3\textwidth}
    \includegraphics[width=0.28\textwidth,trim={0 2cm 2cm 2cm}]{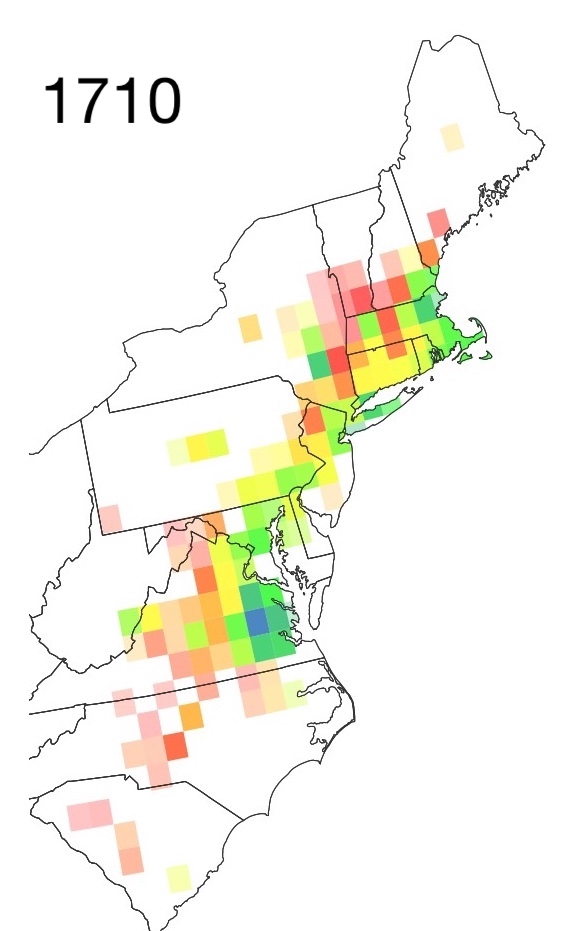}
\end{wrapfigure} 
\pagebreak
Birth-death symmetry S is a more information-rich descriptor for the frontier line.  In our maps, profiles are assigned to locations at age 20 (SM1, SM5), thus areas with negative mean symmetry (orange and red in the maps) are a view to the future, where these people died $\sim$20-50 years after the map date.  The map for 1710 (right) is nearly three generations before the first 1790 census data.  We see the frontier line in yellow (i.e. $S \approx 0$) as well as the future of those crossing that line 1710-1740, namely settlement into Vermont, New Hampshire, eastern New York, and movement west from the Tidewater region.

The symmetry map for 1830, below, bears comparison to the Gannett and $f_{stay}$ maps above.  Again the $S \approx 0$ boundary corresponds to the traditional frontier line, consistent with the expected transition between dispersive and concentrative migration.  The future-view clearly shows these migrants’ destinations, including the beginnings of western urbanization in Chicago, Salt Lake City, San Francisco, and Los Angeles.

Complete 1620-1950 maps for these and other metrics are at \url{http://scaledinnovation.com/gg/migration/migration.html} .

\begin{figure}[H]
    \includegraphics*[width=12cm]{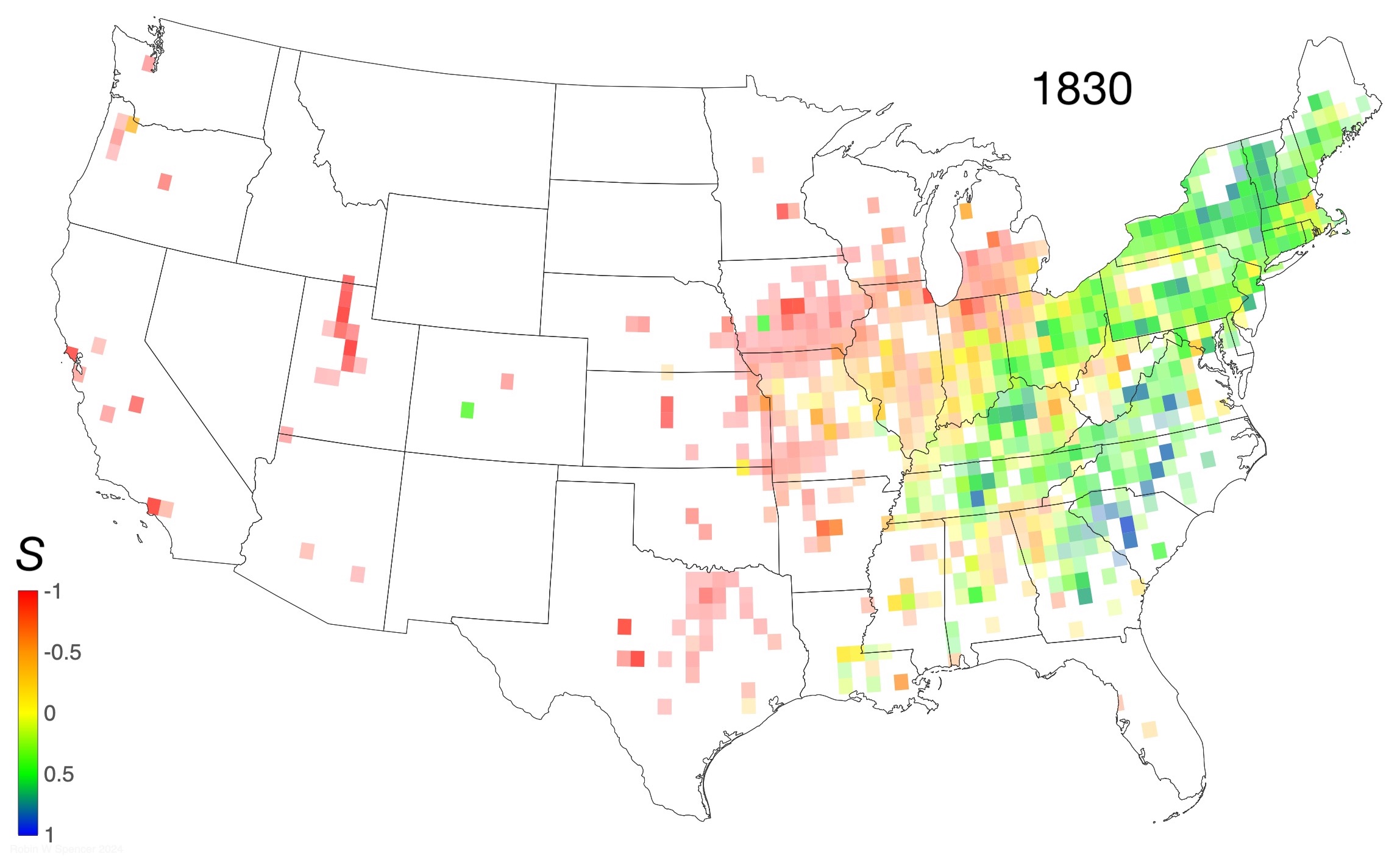}
    \centering
\end{figure}

\end{document}